\documentclass{natureTest}

\usepackage{xr}
\usepackage{subcaption}
\usepackage{graphicx}
\usepackage{amsmath }

\graphicspath{{Figures/}}

\title{LOOPER: Inferring computational algorithms enacted by neuronal population dynamics.}

\author{Connor Brennan$^{1}$, \& Alex Proekt$^2$}

\begin{document}

\maketitle

\begin{affiliations}
	\item Department of Neuroscience, University of Pennsylvania
	\item Department of Anesthesiology and Critical Care, University of Pennsylvania
\end{affiliations}

\begin{abstract}
	Recording simultaneous activity of hundreds of neurons is now possible. Existing methods can model such population activity, but do not directly reveal the computations used by the brain. We present a fully unsupervised method that models neuronal activity and reveals the computational strategy. The method constructs a topological model of neuronal dynamics consisting of interconnected loops. Transitions between loops mark computationally-salient decisions. We accurately model activation of 100s of neurons in the primate cortex during a working memory task. Dynamics of a recurrent neural network (RNN) trained on the same task are topologically identical suggesting that a similar computational strategy is used. The RNN trained on a modified dataset, however, reveals a different topology. This topology predicts specific novel stimuli that consistently elicit incorrect responses with near perfect accuracy. Thus, our methodology yields  a quantitative model of neuronal activity and reveals the computational strategy used to solve the task.
\end{abstract}

\section*{Main}

Dynamics of even simple nervous systems arise as a result of nonlinear interactions among many component processes. Despite this apparent complexity, behaviorally-relevant neuronal activity observed during performance of behavioral tasks can often be modeled using only a few parameters\cite{Afshar2011,Carnevale2015,Laurent2001,churchland2012neural,Harvey2012,Kaufman2014,Kobak2016,Mante2013,Pandarinath2015,Sadtler2014}. This implies that high-dimensional neuronal activity is driven by low-dimensional neuronal dynamics\cite{Gao2017,Pandarinath2018,churchland2012neural,Mante2013,Kato2015,Williams2020,Gao2015}. However, the relationship between these low-dimensional neuronal dynamics and the computations used by the nervous system to solve the behavioral task is not clear.

 One way to improve our understanding of the relationship between neuronal dynamics and computation is to compare the dynamics observed in the brain and in artificial neuronal networks trained on a similar task.  Artificial networks are becoming an increasingly important model system  in neuroscience\cite{Simonyan2015,Vinyals2019,Senior2020, Barak2017}. Patterns of activity similar to those in the brain have been observed in artificial neural networks trained on similar tasks\cite{Cueva2018,Mark,Yamins2016,Banino2018}. Recurrent neural networks (RNNs) in particular, have another important function as a model  of how adaptive behavior could emerge in an abstract dynamical systems. Experiments in RNNs reveal that while the exact patterns of activity during execution of a task depend strongly on the specific components that make up the system, the topological description remains invariant so long as task demands are the same\cite{Maheswaranathan2019}. A similar observation has also recently been made in the nervous system of nematode \textit{C. elegans}\cite{Brennan2017,Brennan2019a}. This suggests that while the differences in the physical implementation do impose system-specific constraints reflected in the different patterns of neuronal activity, the more abstract topological description of the dynamics is constrained by task-specific requirements and is closely related to the computation \textit{per se}.

Uncovering this topological description on the basis of experimental observations therefore is an essential step in understanding the computational strategies used by the brain and other intelligent systems. Several modelling techniques\cite{Pandarinath2018,Raghu,Kornblith2019} are remarkably successful at modeling neuronal activity. However, these techniques are optimized to recover the geometric rather than topological structure\cite{Maheswaranathan2019} and therefore do not readily reveal the computational strategy used by the system to solve the task. 

Here we present LOOPER -- an exploratory analysis method for non-parametric and unsupervised discovery of behaviorally salient neuronal dynamics from high-dimensional, noisy, and non-linear neuronal activity data. LOOPER simultaneously provides both a quantitative model and qualitative understanding of the underlying dynamics. The quantitative model allows for precise predictions of future neuronal activity. The primary advancement of LOOPER is that it also captures a qualitative topological description of the dynamics -- the computational scaffold\cite{Brennan2017,Brennan2019a,Maheswaranathan2019}. To arrive at the computational scaffold, dynamics are succinctly approximated by a collection of ''loops''. We define a ''loop'' to be a recurrent trajectory traced out by the activity of the nervous system during repeated executions of similar computations. The branching patterns of these loops define the topology and are closely related to the computational scaffold -- judgments made by the system during task performance. This topological definition of a loop differs significantly from the classical definition of an oscillation, as the phase progression along the loop need not be constant. Furthermore, the loop may assume an arbitrary shape so long as similar trajectories are observed repeatedly. We show that loops are the ideal dynamical structures for reliably storing behaviorally relevant information in noisy dynamical systems. Therefore, the unsupervised loop extraction reflects the computational requirements of the task in an unbiased fashion.

The successful application of LOOPER relies on several parsimonious assumptions. First, the dynamics must not change over the course of the experiment (i.e. no learning can occur). This standard assumption may be implemented in a piecewise fashion by using short recording sessions during learning. We also assume that at least some aspects of the recorded signals are behaviorally-relevant. It is not necessary to record all the salient dynamical variables. Nor is it critical that these variables are related in a simple fashion to neuronal activity. So long as the salient variables can be recovered from experimental observations, LOOPER will identify recurrent trajectories. If there are additional driving forces (such as task irrelevant dynamics) these may need to be removed by averaging over trials or machine learning-based approaches\cite{Pandarinath2018}. 

The most critical assumption that enables LOOPER to extract behaviorally meaningful dynamics is that the system is adapted to perform a task. Such adaptation allows significant simplification of dynamics in noisy systems such as the brain. Adaptive task performance suggests that circumstances considered to be similar by the system elicit similar dynamics. Thus, neuronal activity trajectories elicited under similar circumstances will tangle under the influence of stochastic forces. If trajectories starting from nearby points tangle, then the differences in the initial conditions cannot carry behaviorally salient information. Conversely, if nearby trajectories diverge, then the differences in initial conditions are behaviorally salient. Thus, a set of tangled trajectories can be safely approximated by a single recurrent trajectory bundle -- a loop. We will show that while this approximation offers tremendous simplification, it does not result in appreciable loss of information. 

We apply LOOPER to neuronal activity recorded from rhesus macaques during a working memory task. LOOPER reconstructs the observed neuronal activity with a mean correlation of 0.97 $\pm$ 0.003 (mean $\pm$ standard deviation). We use the model to simulate predicted neuronal activity and obtain a mean correlation of 0.87 $\pm$ 0.07 (mean $\pm$ standard deviation). Further, we show that dynamics in a recurrent neural network (RNN) trained to perform the same working memory are topologically identical. However, if the RNN is trained on a modified set of inputs, it exploits a weakness in the training set to arrive at comparable task performance using a distinct computational scaffold. LOOPER uncovers this scaffold in a fully unsupervised fashion. We validate that we have indeed uncovered the computational scaffold and predict with nearly 100\% accuracy specific novel combinations of inputs that lead to consistently incorrect decisions, while other novel combinations of inputs yield correct results. Thus, LOOPER is capable of extracting the computations being performed by a nonlinear stochastic dynamical system in a fully unsupervised fashion.

\section*{Results}
\subsection{Overview of LOOPER.}

Neuronal dynamics can be described by a stochastic dynamical system of the form:
\begin{equation} \label{Eq:1}
\dot{\mathbf{x}}(t) = \mathbf{F}(\mathbf{x}(t),\mathbf{u}(t)) + \mathbf{\eta}(t).
\end{equation} 
The state of the system $\mathbf{x}(t)$ evolves in time according to some nonlinear function $\mathbf{F}$ and a stochastic noise term $\mathbf{\eta}(t)$. $\mathbf{F}$ can additionally accept inputs, $\mathbf{u}(t)$. Due to the stochastic nature of such systems it is not possible to precisely model the time-evolution of any one trajectory starting from a particular location in phase space, $\mathbf{x}(t)$. Instead, we model the time evolution of the probability density of many instances of the system, $P(t)$, in a local neighborhood around $\mathbf{x}(t)$. The time evolution of $P(t)$ is governed by the Fokker-Planck equation.
In most situations of interest  $\mathbf{F}$ in  Equation~\ref{Eq:1} is unknown or is too complex to be analytically or computationally tractable. Thus, we  approximate the Fokker-Planck equation using Markov models. 

Diffusion maps\cite{nadler2006diffusion,Yair2017,coifman2006diffusion} are a natural choice for the construction of these Markov models. There is a clear connection between diffusion maps and the discrete Fokker-Planck equation\cite{nadler2006diffusion}. The basic idea behind diffusion maps is to break up the system into small neighborhoods centered around each experimental observation. First, the probability density $P(t)$ in the neighborhood of each experimental observation, $\mathbf{x}(t)$, is estimated using a Gaussian kernel. $P(t)$ expresses the probability that the system starting out at $\mathbf{x}(t)$ diffuses to a different experimental observation within the local neighborhood in one time step. As the system is allowed to evolve in time through the exponentiation of the diffusion map, the global nonlinear manifold is stitched together from the locally Euclidean neighborhoods. In this way diffusion maps provide nonlinear dimensionality reduction that preserves topological information. Preservation of the topological information then allows LOOPER to simplify the resulting diffusion map as a collection of interconnected loops (Methods). The pattern of connections between loops defines the topology of the manifold and is closely related to the computations performed by the system to solve the task. 

To introduce LOOPER we construct a 2-dimensional stochastic dynamical system (Figure~\ref{Fig:1}). Dynamics similar to those illustrated in Figure~\ref{Fig:1} have been observed in the central nervous system of nematode \textit{C. elegans}\cite{Brennan2017,Brennan2019a,Kato2015}. As long as the system traverses one of the loops, one kind of locomotion behavior is performed. Switching between different loops is associated with switches in locomotor behavior. Thus, the loop structure reflects the global command structure of the animal's locomotive behavior. For the sake of illustration in Figure~\ref{Fig:1} we simulate similar dynamics and develop an intuition for why such loop structures arise in diverse systems and tasks. 

The dynamics of stochastic dynamical systems are given by two essential ingredients: the energy landscape (Figure~\ref{Fig:1}B, heat map) and the flux (Figure~\ref{Fig:1}B, thick gray arrows). The only possibility that admits fluxes and is consistent with the stationarity assumption is that all of the fluxes form divergence free and topologically closed loops\cite{Wang2008} (Supplementary). The primary objective of LOOPER is to discover these loops solely from the observations of the system in an unsupervised fashion. 

In the absence of noise, trajectories spanned by the dynamical system never cross. Noise will cause trajectories to tangle over time. The moment two trajectories cross, information concerning all of the events prior to the crossing is erased. Thus, information carrying capacity of a stochastic dynamical systems is drastically limited. Specifically, the only information that reliably survives over time is the identity of the trajectory bundle consisting of many tangled system trajectories 
(Supplementary, Figure~\ref{Fig:noise}). LOOPER uses this intuition to combine trajectories into a bundle if they evolve in a coherent fashion through phase space. 

To see why such trajectory clustering yields a meaningful simplified description of the system, consider two sets of points (red and blue) starting from a relatively small neighborhood in phase space (Figure~\ref{Fig:1}C). While the points are initially close to each other, the trajectories do not merge over time because the initial conditions belong to two different bundles. Thus, information about the starting position is preserved. Conversely, trajectories starting from a similar sized neighborhood in a different location in phase space merge (Figure~\ref{Fig:1}D) because both sets of initial conditions belong to the same bundle. In this case, information about differences in initial conditions is quickly erased. This implies that the complex dynamics of a system such as that shown in Figure~\ref{Fig:1} can be approximated by modeling the evolution of the system as a function of the identity of the loop  formed by a set of tangled system trajectories and the phase along it without significant loss of behavioral information. This is why the primary objective of LOOPER is to identify distinct loops in the dynamics of the system. 

LOOPER builds its model of dynamics using several critical modifications to diffusion maps\cite{Brennan2017,Brennan2019a,nadler2006diffusion,coifman2006diffusion} (Methods). One key modification is to allow for anisotropic deformation of the local neighborhoods. Note that the initially isotropic point cloud is preferentially stretched along the local direction of the loop and shortened in all directions orthogonal to it (red point clouds in Figure~\ref{Fig:1}B). Correction for anisotropy is implemented by rescaling each local neighborhood (Methods). 

The second modification of the diffusion maps is the construction of an asymmetric diffusion distance. This modification assures that the most likely transitions occur on the one off diagonal (i.e. any given state is most likely to transition to the next observed state). After appropriate normalization, this modified diffusion map (Figure~\ref{Fig:diffusionMap}) is a discrete approximation of the Fokker-Planck operator which allows for both diffusion and cyclic flux. Given relatively short observations, a high dimensional system is unlikely to pass through the same exact point twice. We therefore make use of the Fokker-Planck approximation to determine if two trajectories pass through similar point clouds. Specifically, to determine if the trajectories starting from states $\mathbf{x(i)}$ and $\mathbf{x(j)}$ can be safely combined into the same trajectory bundle, we compute the similarity between the point clouds observed after some time $i+\Delta t$ and $j+\Delta t$ (Methods, Figure~\ref{Fig:clusteredMap}). This key modification allows us to abstract away from the specific features of neuronal trajectories and arrive at the topological description of the dynamics\cite{Brennan2017,Brennan2019a} (Methods, Figure~\ref{Fig:loopClusters}). 

The specific implementation of the clustering used by LOOPER is described in the Methods Section. The method proceeds in a step-by-step fashion to facilitate parameter exploration at each step. Parameters used to generate the figures in this text are listed in Table~\ref{Tab:parameters}. Parameter robustness is explored in Figure~\ref{Fig:parameterBreakdown}. Once loop clustering is performed, a model of the dynamics is constructed by assigning points in each loop to phase bins (Methods) and averaging the location of all experimental observations in the bin. The resulting model can be succinctly summarized as a manifold of interlocking loops, each corresponding to a behaviorally relevant trajectory bundle (Figure~\ref{Fig:1}E, left). Each discrete location in phase space is uniquely specified by its loop identity and the phase along the loop (Figure~\ref{Fig:1}E, right). The modified diffusion map then is used to estimate $\mathbf{F(x)}$ at each location defined by the identity of the loop and the phase along it. Thus, the manifold is effectively a 2-dimensional object. This means that even though the model itself has many degrees of freedom, the resulting Markov matrix tends to be quite sparse (Figure~\ref{Fig:1}E, middle). This approach works even if the stochastic dynamical system exhibits chaotic behavior (Figure~\ref{Fig:Lorenz}). As we will show below, while this tremendous simplification greatly improves the interpretability of the dynamics, it does not significantly degrade the ability of the model to accurately simulate neuronal dynamics.

\clearpage

\begin{figure}
	\centering
	\includegraphics{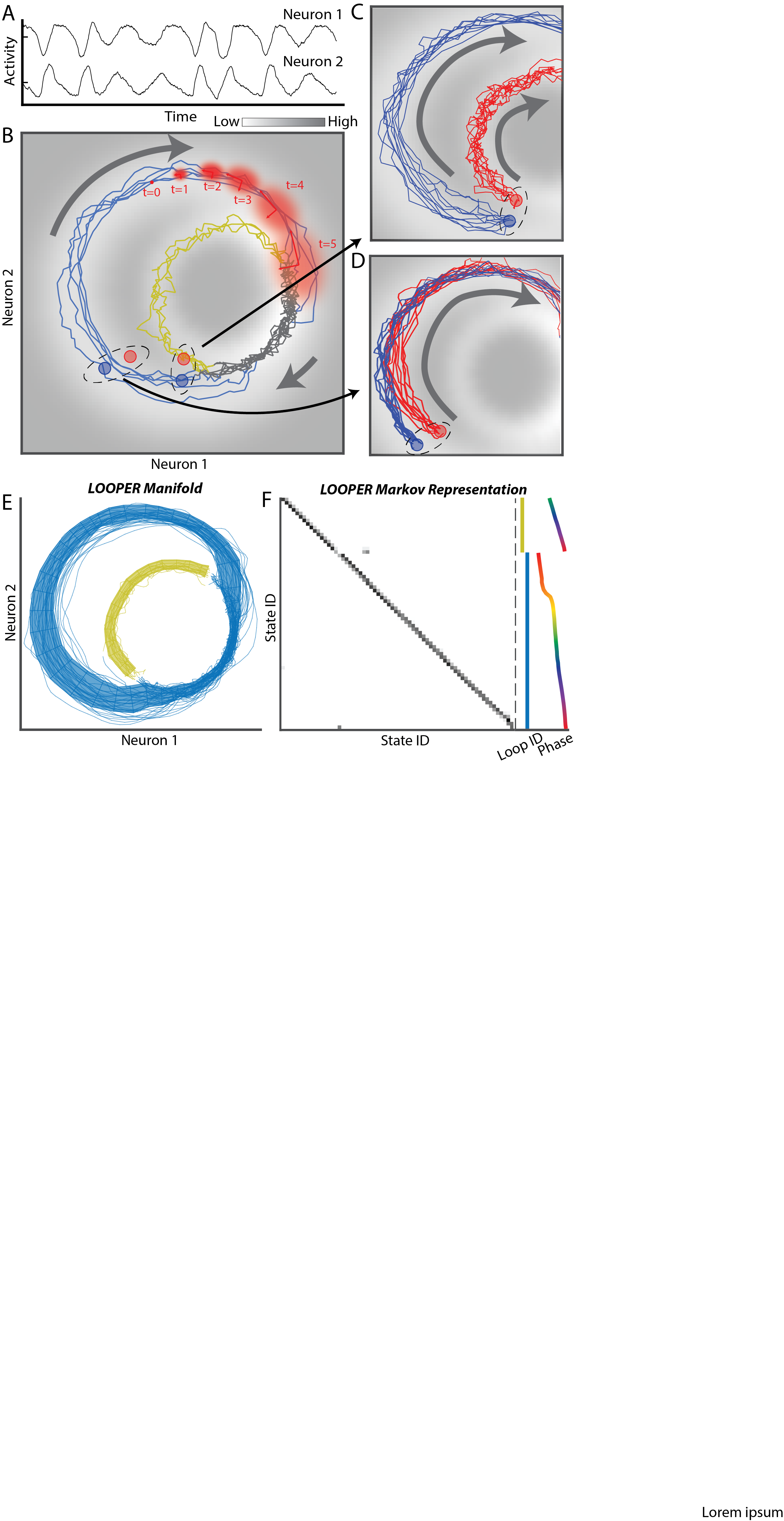}
	\phantomcaption
\end{figure}
\begin{figure}
	\ContinuedFloat
	\caption{{\textbf{LOOPER model deconstructs a stochastic dynamical system into loops}
			\textbf{A)} Simulated activity of two neurons constructed by integrating equations of motion defined by the gradient of a potential function and cyclic flux. \textbf{B)} Activity of the neurons in \textbf{A} is plotted against each other. This plot makes up the entire phase space. For analysis of experimental data, phase space must be inferred from experimental measurements (Methods). Heat map (gray) shows the energy potential. Direction and magnitude of local flux is shown by the gray arrows. Trajectories, colored according to the identity of the trajectory bundle assigned by LOOPER, correspond to repeated activity patterns of the system. The classification of the  region where two bundles merge is arbitrary and is colored in gray. Red clouds show the anisotropic deformation of the the probability density starting from a single local neighborhood at $t=0$ as the system evolves in time. \textbf{C)} 10 simulated trajectories starting from two local neighborhoods (red and blue circles) in \textbf{B}. Grey arrow shows the direction of flux. Trajectories diverge over time because they belong to two distinct bundles. \textbf{D)} Same as \textbf{C} with the exception that in this case the 10 trajectories starting from either the red or blue neighborhood in \textbf{B} merge because they belong to the same bundle. \textbf{E)} LOOPER manifold built on the data in \textbf{A}. The simulated trajectory is colored by loop identity. Shaded regions denote 1 standard deviation from the mean of each phase bin along the manifold. \textbf{F)} Each state of the system (activity of neurons 1 and 2 in this case) is assigned a unique combination of phase bin and loop identity (right). LOOPER constructs a Markov model of transitions between these states (left). Most transitions occur on the one-off diagonal indicating flow along the loop. The few exceptions correspond to transitions between the two loops. Note that the phase velocity is not always constant along the loop (right, elbow in phase). Thus, LOOPER does not rely on oscillations in order to identify loops in system dynamics.}
		\label{Fig:1}}
\end{figure}

\clearpage

\subsection{LOOPER model reconstructs observed neuronal activity.}
We first demonstrate the capabilities of LOOPER by applying it to previously published neuronal activity data from a working memory task in rhesus macaques\cite{Kobak2016,Romo1999}. A schematic of the task is shown in Figure~\ref{Fig:2}A. The monkey is trained to compare the frequency of two stimuli -- vibrations presented via fingertip. The reference stimulus (F1) and the comparison stimulus (F2) are separated by a 3s time interval. The monkey reported whether the frequency of F1 was higher or lower than F2 by pressing one of two buttons. Neuronal spiking was recorded from prefrontal cortex (Figure~\ref{Fig:2}B) and converted to firing rates (Figure~\ref{Fig:2}C). The dataset is made up of hundreds of disjoint recording sessions each capturing between 1 and 7 individual spiking units. Thus, as in previous work on these data\cite{Kobak2016}, we make use of pseudotrials to study the neuronal population dynamics (Methods). Note that although this preprocessing step is supervised in a sense that it necessarily uses task information, LOOPER is not given any information about the identity of F1 or F2. We focused on the 6 stimulus combinations with the most observed trials. This yielded a subset of 179 neurons with at least 10 trials for each of the 6 stimulus combinations. The average success rate of the monkeys on this task is 95\%. Thus, the neuronal networks that give rise to the task performance can be said to be well-adapted to the task. Only the correct responses were considered in the model due to the paucity of errors.   

To construct the model of the dynamics we reconstruct phase space using delay embedding applied to the neuronal activity projected onto the first 10 principal components (PCs) (92.5\% variance explained). Trajectories spanned by the system in this reconstructed phase space are clustered  using LOOPER (Methods). To validate the model, the LOOPER manifold can be mapped back into the original raw neuronal firing space by averaging the neuronal activity in a given bin of the manifold. This back projection from the manifold to the activity space is shown in Figure~\ref{Fig:2}D. The across-trial mean correlation coefficient for all 60 trials (179 neurons; 10 trials for each condition) vs their corresponding condition-averaged neuronal activity is 0.97 $\pm$ 0.003 (mean $\pm$ standard deviation). 

LOOPER does not just recapitulate neuronal activity. It is a model of neuronal dynamics that generates activity patterns \textit{de novo}. To further validate the model, we start the LOOPER simulation from a snapshot of neuronal activity observed at some time $t$ during the task and predict subsequently observed neuronal firing rate. The accuracy of this prediction for three example neurons is shown in (Figure~\ref{Fig:2}E). Figure~\ref{Fig:2}F shows the distribution of correlation coefficients computed for all simulated trials. The across-trial mean correlation coefficient for the predictions vs their corresponding condition-averaged neuronal activity is 0.87 $\pm$ 0.07 (mean $\pm$ standard deviation). Therefore, LOOPER model yields quantitatively accurate reconstruction and prediction of neuronal activity.

\clearpage

\begin{figure}
	\centering
	\includegraphics{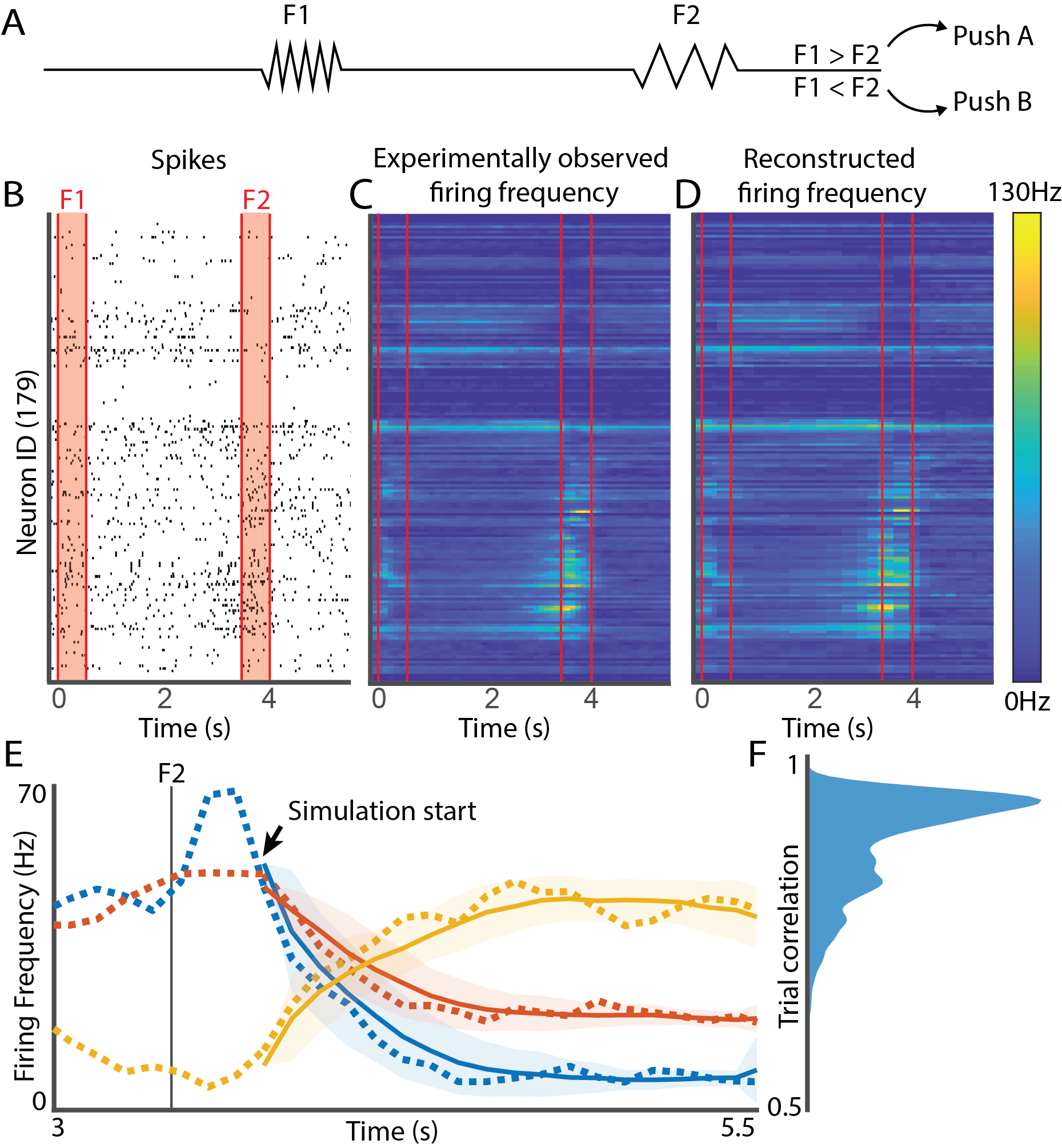}
	\phantomcaption
\end{figure}
\begin{figure}
	\ContinuedFloat
	\caption{\textbf{LOOPER reconstructs neuronal activity during a working memory task}
			\textbf{A)} Schematic of the working memory task. The monkey receives a sequence of two vibrational stimuli applied to the fingertip (F1 and F2), with an interstimulus delay of 3s. The monkey must push one of the two buttons depending on whether the frequency of F1 is greater than F2, or not. We use only combinations of F1 and F2  that had more than 10 recordings for each neuron. \textbf{B)} Spike raster of a single pseudotrial constructed from aligning multiple recording sessions for a particular combination of F1 and F2. Red lines mark on the onset and offset of F1 and F2. \textbf{C)} Average instantaneous firing frequency for the combination of F1 and F2  shown in \textbf{B}. Firing frequency was calculated by convolving the spike raster with a 30ms Gaussian kernel. \textbf{D)} Reconstruction of the average firing rate from the LOOPER model ($R^2$ = 0.97 $\pm$ 0.003, mean $\pm$ standard deviation). 
			\textbf{E)} Simulations of 3 representative neurons starting from a state corresponding to the offset of F2.
			Dotted lines show condition-averaged observed firing rate of a single representative neuron. Solid line and confidence interval show the mean simulated firing rate and one standard deviation. The simulation is run for 1.9 seconds (19 time steps). These simulations are repeated for each neuron (179) in all of the 6 task conditions starting from the same time point (4.0s after F1, black arrow). Simulated neuronal firing correlates to condition-averaged activity with $R^2 = 0.87 \pm 0.07$ (mean $\pm$ standard deviation). \textbf{F)} Smoothed histogram of trial correlations. Each of the 6 conditions is simulated 100 times and compared to corresponding condition-averaged neuronal activity.}
		\label{Fig:2}
\end{figure}

\clearpage

\subsection{LOOPER captures  distinctive dynamics of recurrent neural networks.}
There is considerable interest in directly comparing neuronal activity in biological brains and artificial recurrent neural networks (RNNs)\cite{Barak2017}. RNNs have been shown to reproduce spontaneous neuronal activity\cite{van1996chaos}, activity in the prefrontal cortex during tasks involving temporal integration of sensory information\cite{Mante2013}, memory formation and sequence generation in the posterior parietal cortex\cite{rajan2016recurrent}, muscle control\cite{sussillo2015neural},  and  higher cognitive functions\cite{enel2016reservoir, Cueva2018}. More generally, however, both the biological and artificial neuronal networks can be thought of as dynamical systems that have been adapted to perform a task. This suggests that dynamics in biological and artificial neuronal networks trained to perform the same task should exhibit some similarities.  

Thus, to compare the dynamics that arise in the primate prefrontal cortex to RNNs, we trained a RNN to solve the same working memory task (Methods). A schematic of the RNN composed of 100 long-short term memory (LSTM) units is shown in Figure~\ref{Fig:3}A. In order to mimic real biological systems in which neuronal firing is intrinsically noisy, we add noise to each  LSTM and to the input stimulus (Methods). In the monkey data we had to create pseudotrials because only a handful of individual neurons were recorded in each experimental session. Thus, information about stimuli, while not directly accessible to LOOPER, was inadvertently introduced. In contrast, the reconstruction of the LOOPER manifold based on the RNN data is fully unsupervised. The hidden variable activity for all 100 LSTM units are shown in Figure~\ref{Fig:3}B. We again use LOOPER to build a manifold of the dynamics and back project the manifold into activity space as in Figure~\ref{Fig:2}D. The across-trial mean correlation coefficient (60 trials, 200 hidden variables; 10 trials for each condition) of the reconstructed activity (Figure~\ref{Fig:3}C) vs their corresponding condition-averaged neuronal activity is 0.99 $\pm$ 0.007 (mean $\pm$ standard deviation). This illustrates that LOOPER is able to recover a low dimensional model from high dimensional noisy data in a fully unsupervised fashion.


The LOOPER manifolds extracted from the primate (left) and RNN (right) activity are shown in  Figure~\ref{Fig:3}D. These manifolds consist of interconnected trajectory bundles or loops. Note that shapes of the trajectory bundles identified by LOOPER in both biological and artificial networks are rather complex. Nonetheless, LOOPER is able to quantitatively reconstruct the neuronal activities in both systems. Furthermore, both the shape of the trajectory bundles and the neuronal activity differs significantly between the two systems. The separation of the trajectory bundles in the prefrontal cortex can be better visualized when projected onto demixed PCA in Figure~\ref{Fig:geometry}. This supervised trajectory separation plays no role in model construction and is used purely for visualization purposes here.  

The apparent differences in neuronal activity and the shape of the trajectory bundles in the RNN and the brain may suggest that fundamentally different dynamics are engaged to solve the task in the two systems. These differences, however,  are limited to the geometry of the manifold. Using the simplified model of the dynamics constructed by LOOPER we will show in Figure~\ref{Fig:4} that the topology of the manifolds is identical in both systems.

\clearpage

\begin{figure}
	\centering
	\includegraphics{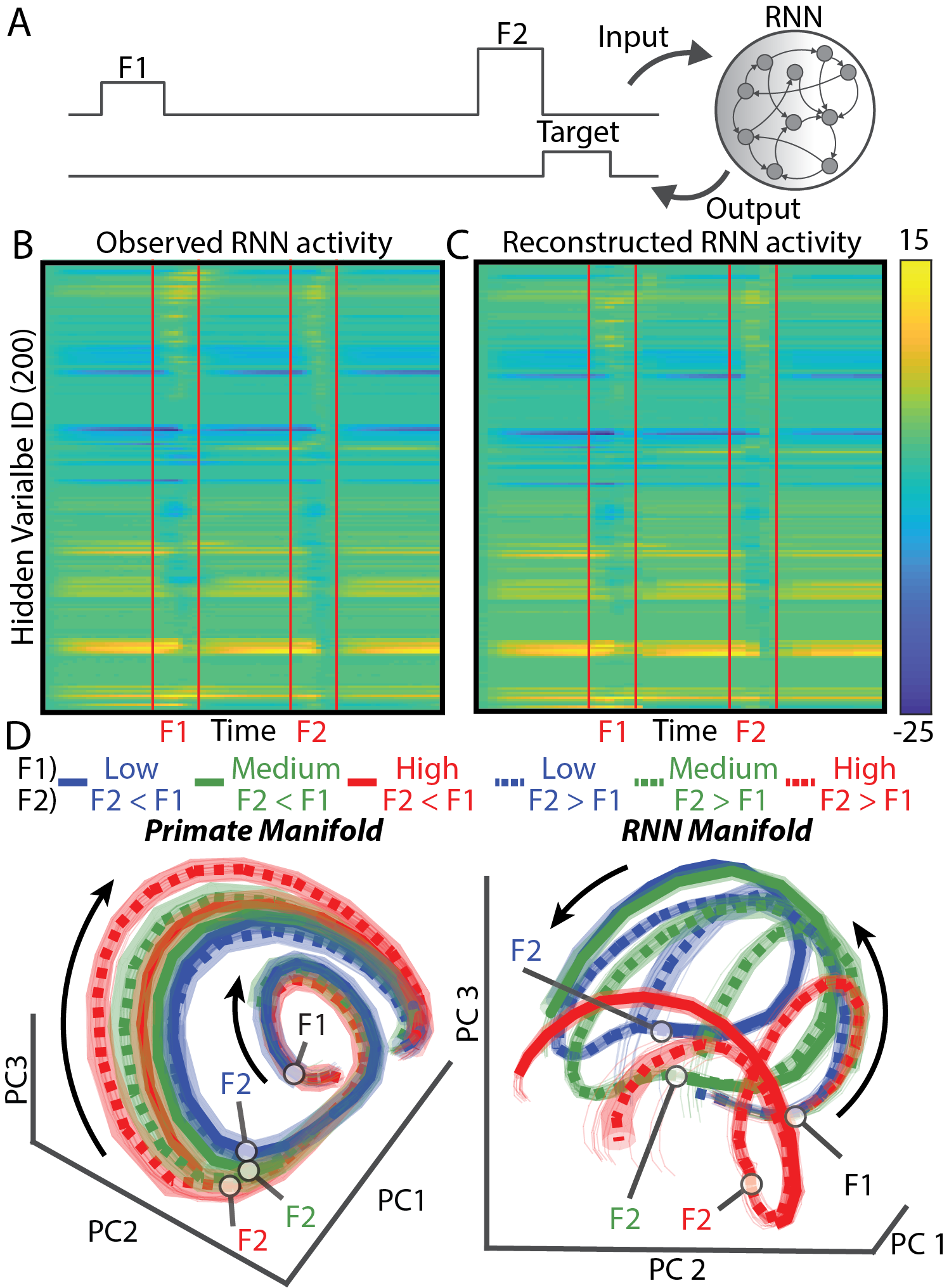}
	\phantomcaption
\end{figure}
\begin{figure}
	\ContinuedFloat
	\caption{\textbf{Fully unsupervised extraction of neuronal activity from a RNN.}
		\textbf{A)} RNN version of working memory task. The RNN is fed a scalar value representing the frequency of F1 and F2 stimuli. This value is corrupted by noise. The RNN must learn to output a target value corresponding to whether F1 is greater than or less than F2 only during the period of time directly following the offset of F2. \textbf{B)} Average activity in the hidden variables of the network for a particular set of stimuli. Each of the 100 LSTM units has both a hidden state ($h$) and a cell state ($c$) for a total of 200 hidden variables. Red lines mark the onset and offset of F1 and F2. \textbf{C)} Reconstruction of the average activity from the LOOPER model ($R^2$ = .99 $\pm$ 0.007, mean $\pm$ standard deviation). For the purposes of reconstruction, each time point is set to the mean activity of its assigned bin. \textbf{D)} Manifolds of dynamics constructed by LOOPER (left $\rightarrow$ primate, right $\rightarrow$ RNN). Thin trajectories mark a single trial colored by F1. Thick trajectories show the average position of each phase bin. Dashed lines indicate F2 $>$ F1 and solid lines indicate F2 $<$ F1. Position and variance of each bin is computed by assigning each data point a phase bin and then taking the mean and standard deviation of each bin. Shaded areas shows 1 standard deviation from the mean (averaged across the 3 PCs used for plotting). The F1 and F2 stimulus timings are shown as white markers on the manifolds. Black arrows show phase velocity along the manifolds. For visualization purposes, the binned data are projected onto the first three principal components.}
	\label{Fig:3}
\end{figure}

\clearpage

\subsection{Computational scaffold extracted by LOOPER is the same in both primate and RNN.}

To solve the task in Figure~\ref{Fig:2} and Figure~\ref{Fig:3}, a system must first classify the initial stimulus (F1) and remember its value. Then the second stimulus (F2)  must be classified  and compared to the remembered reference stimulus. The outcome of this comparison is indicated by a button press (primate) or output class (RNN). A schematic of the computations needed to solve the task is shown in Figure~\ref{Fig:4}A.

To uncover the computational scaffold actually used by the brain and the RNN, each trial is mapped onto its corresponding loop identity at each time point (Figure~\ref{Fig:4}B and C). This corresponds to re-plotting the data from Figure~\ref{Fig:3}D as a function of task events rather than as a function of position in phase space. This reveals that unique stimulus combinations cluster nicely into appropriate loops. This loop structure maps in a one to one fashion to the different steps involved in task performance. The system starts off as a single point (start state). It then receives the first stimulus (F1) and quickly diverges into 3 distinct bundles corresponding to each F1 stimulus value (classifying F1, primate validation accuracy 99\%, primate p-value $<$ 0.0001 bootstrap vs. null model). The bundles remain distinct for the duration of the interstimulus delay (remembering F1, primate validation accuracy 98\%, primate p-value $<$ 0.0001 bootstrap vs null model). Upon receiving F2 each of these 3 distinct bundles further bifurcates into two -- one branch representing F1 $<$ F2 and the other representing F1 $>$ F2 (comparing F2 to the remembered F1, primate validation accuracy 98\%, primate p-value $<$ 0.0001 bootstrap vs null model). Note that this bifurcation does not simply encode F2. The same F2 value yields distinct trajectories when encountered after different F1s. Thus, the dynamics depend not just on the stimulus, but also on the state of the system. In the context of this particular task, the state of the system encodes the remembered F1 stimulus. This hierarchy of dependencies in the task demands gives rise to the specific topological structure of the manifold. Finally, after the system produces a response, the 6 distinct loops representing each of the 6 conditions begin to converge (task end). This convergence reflects the fact that the information about F1 and F2 is no longer salient.

The manifold in Figure~\ref{Fig:4}B was found using a subset of data. We validate the manifold by projecting the remaining subset of data not used in manifold construction onto the LOOPER manifold (Methods, Figure~\ref{Fig:validation}). The correct loop identity can be decoded 96\% of the time (n=3960, p-value $<$ 0.0001 bootstrap vs null model). Thus, the topological structure discovered by LOOPER is generalizable across the experimental dataset and therefore the computational scaffold is consistent across trials. Indeed, LOOPER extracts the same computational scaffold from the dynamics of an RNN trained to solve the same working memory task. This implies that, while the geometric description of the dynamics may be distinct (Figure~\ref{Fig:3}D), the topological description is conserved both across trials and across distinct artificial and biological systems.

\clearpage

\begin{figure}
	\centering
	\includegraphics{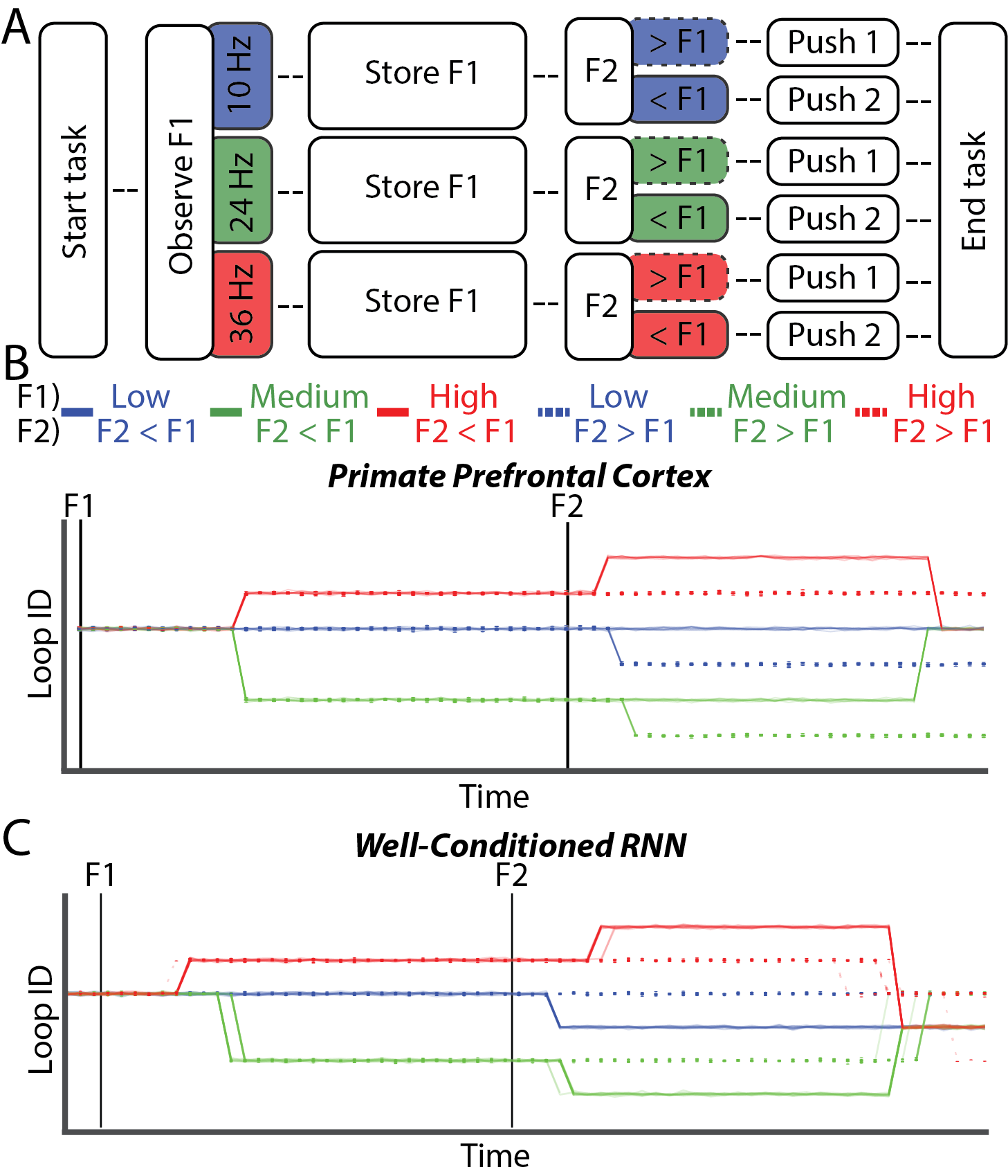}
	\phantomcaption
\end{figure}
\begin{figure}
	\ContinuedFloat
	\caption{\textbf{Computational scaffold of the working memory task is correctly extracted by LOOPER from neuronal recordings and artificial networks}
			\textbf{A)} A schematic of the computational requirements of the working memory task. Each  box corresponds to a required action. Each dotted line shows the number of distinct pieces of information that action imposes on the system. Colors correspond to the specific stimuli combinations observed in the dataset. Note that the observation of F1 causes the system to diverge into 3 distinct trajectories, and the observation of F2 causes each of those trajectories to branch into another 2 trajectories. \textbf{B)} Computational scaffold extracted from the primate data by LOOPER. The same branch structure is observed as in \textbf{A}. The loop IDs assigned by LOOPER are plotted as a function of time and colored by each stimulus combination (shown above \textbf{B}). Similar trajectories are grouped together, illustrating which pieces of information are relevant at any given time.  \textbf{C)} When trained on a similar task, the RNN recapitulates the computational scaffold observed in the monkey (Figure~\ref{Fig:3}B). }
		\label{Fig:4}
\end{figure}

\clearpage

\subsection{The computational scaffold extracted by LOOPER predicts task performance on novel stimuli.}

Special consideration must be given to the data used to train the RNN. To obtain the scaffold in Figure~\ref{Fig:4}C, the RNN was trained on 3 unique F1s and 4 unique F2s. The key aspect of the training set is that the same F2 can be presented after different F1s such that the correct response depends on both F1 and F2. Interestingly, this was not the case for the monkey experiment. In the primate data some F2 stimuli were only presented after a single F1. Thus, strictly speaking, it is possible to correctly solve the task in some instances without any information concerning F1 (Figure~\ref{Fig:5}A). For instance, if F2 happens to be 25, the correct answer is always “greater than” regardless of F1. Similarly, when F2 is 30 the response is always "less than". When we trained an RNN on a similar set of stimuli as used in the monkey experiment, the LOOPER model (Figure~\ref{Fig:5}A) extracted a topologically distinct manifold. Specifically, the 2 loops corresponding to the F1 stimuli 20 and 40 are fused before the onset of F2. The other loop unequivocally identifies a specific F1=10 over the entire interstimulus period. This difference in manifold topology could not have been inferred from task performance. The poorly-conditioned RNN has near perfect performance levels on the training sets. 

If the computational scaffold uncovered by LOOPER is correct, then the poorly-conditioned RNN should make specific errors when presented with a novel combination of stimuli. Because, according to LOOPER manifold, the poorly-conditioned RNN does not distinguish between F1=20 and F1=40, it will erroneously output "Greater than" when presented with F1=40 followed by F2=25 (Figure~\ref{Fig:5}B, red arrows).  Similarly, the LOOPER model predicts that F1=20 followed by F2=30 should incorrectly yield response "Less than" in the RNN. Finally, the manifold extracted by LOOPER predicts that the poorly-conditioned RNN will succeed at the the novel stimuli combinations F1=40 followed by F2=15 and F1=20 followed by F2=50. This is exactly what was observed for the poorly-conditioned RNN for both of these stimulus combinations on 99\% of trials (Figure~\ref{Fig:5}C). Simulation of the LOOPER manifold constructed from the poorly-conditioned RNN also gave erroneous results on 100\% of trials (Methods). The failures of the poorly-conditioned RNN is not a reflection of noise in the network. If that were the case, then the outcome on the novel stimulus combination would be variable. In contrast, the LOOPER manifold predicts the specific combinations that lead to consistently wrong answers exactly, while other novel stimulus combinations result in adequate performance.  This suggests that the errors are due to the incorrect algorithm implemented in the RNN dynamics rather than stochastic influences or the generic tendency to overfit the training dataset by the RNN. Indeed, the well conditioned RNN yielded adequate performance on all novel stimulus combinations. Thus, LOOPER reveals that a different computation is being performed in a poorly-conditioned RNN even when the performance level on the training set is identical to that observed in the primate. 

There are fundamental advantages to unsupervised learning of the manifold by LOOPER over supervised techniques. For instance, dPCA of the poorly-conditioned RNN shows that the trajectories reflecting F1=20 and F1=40 are statistically distinct (Figure~\ref{Fig:dpcaTopology}) during the interstimulus interval. This would imply that the poorly-conditioned RNN does indeed encode and remember F1 identity unequivocally. However, this dPCA-based assertion is not consistent with the specific pattern of errors made by the RNN on novel stimulus combinations. LOOPER, in contrast, is able to circumvent this difficulty and correctly combines trajectories into a single bundle. This reveals a fundamental distinction between statistical differences and the behaviorally-salient aspects of neuronal dynamics. Statistical differences in neuronal activity do not in and of themselves imply that the system makes use of these distinctions.

It is remarkable that given a poorly-conditioned training set, the monkey is able to learn the more general computational scaffold, while the RNN solves the problem by simply modeling stimulus statistics. This suggests that although the end result of RNNs and brains may be behaviorally similar on the training set, the learning rules that give rise to successful task performance and therefore the topology of the manifold are fundamentally distinct. Direct comparisons between computational scaffolds observed in biological brains and artificial systems adapted to solve the same task using LOOPER may therefore help elucidate both the similarities and differences in how biological and artificial systems learn the task. 

\clearpage

\begin{figure}
	\centering
	\includegraphics{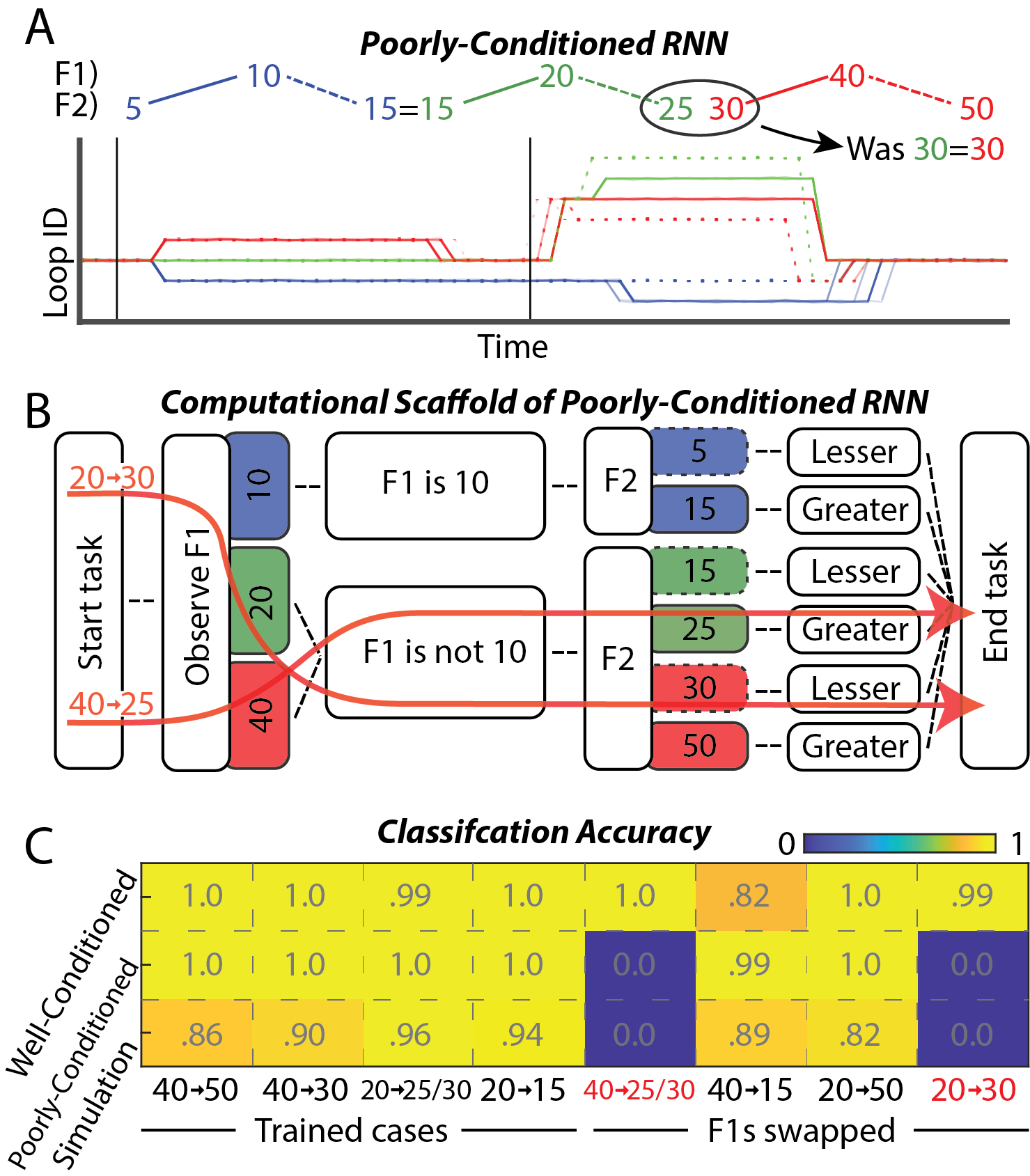}
	\phantomcaption
\end{figure}
\begin{figure}
	\ContinuedFloat
	\caption{{\textbf{LOOPER manifold makes specific predictions about novel stimuli combinations}
			\textbf{A)} When trained on a slightly modified training set the RNN uses a different computational scaffold that takes advantage of a flaw in task design. The key difference is that two F2 stimuli (25 and 30) are only observed after a single F1. In the manifold, the distinction between F1=20 and F1=40 is lost before the onset of F2 because the trajectories corresponding to these values of F1 fuse. \textbf{B)} A schematic of the poorly-conditioned computational scaffold (Compare to Figure~\ref{Fig:4}A). Red arrows show specific predictions about task performance on novel stimuli combinations. The topological model predicts that the RNN does not distinguish F1=20 and F1=40, and relies solely on F2 to perform the task in those cases. Thus, we should be able to swap F1s to get consistently incorrect results. \textbf{C)} Table of classification accuracies for both training and novel stimuli pairs. The well-conditioned network is able to generalize quite well to the novel stimuli (top row). The poorly-conditioned network, however, consistently gives the correct and incorrect answers on the stimuli combinations predicted by LOOPER (middle row). Finally, a LOOPER-based simulation of the poorly-conditioned network also gives the same pattern of errors as the poorly-conditioned network (bottom row).}
		\label{Fig:5}}
\end{figure}

\clearpage

\section*{Discussion}
Here we present LOOPER - a method for extracting loops spanned by the trajectories of a stochastic nonlinear dynamical system solely on the basis of observations of system activity in an unsupervised fashion. We demonstrate the success of this methodology in modeling neuronal dynamics exhibited during a working memory task in a primate and in artificial neural networks. Despite fundamental differences in the activity observed in biological brains and artificial neural networks, the topological structure of the manifolds created in the artificial neural network and in the primate brain is essentially identical. This suggests that it is this topological structure rather than the specific ways in which this structure emerges from the interactions among the component processes that carries the computationally salient information used by the system to solve the task. LOOPER is able to recover this structure solely on the basis of experimental measurements. 

Recordings of hundreds of neurons during task performance is now possible\cite{Kato2015,nichols2017global,Ahrens2013,Ahrens2012,Flusberg2008,Katona2012,Holekamp2008,Grewe2010,Cheng2011,Marre2012,churchland2012neural}. Along with this experimental revolution, novel computational methods for analysis of such complex datasets have been developed. LOOPER makes use of diffusion maps\cite{coifman2006diffusion,nadler2006diffusion}, which fall into a class of nonlinear dimensionality reduction methodologies such as Isomap\cite{tenenbaum2000global}, locally-linear embedding\cite{Roweis2000}, kernel-PCA\cite{Scholkopf1998} and others. The basic approach is to build up the global nonlinear structure in which dynamics unfold from local neighborhoods that can be approximated using linear methods. LOOPER employs several key methodological advances which allowed us to successfully apply diffusion maps to neuronal systems. The first is to account for the anisotropic effect of noise. Similar techniques have been proposed by Singer and Coifman\cite{Singer2009}, and it is likely that many possible implementations could succeed in practical applications. The second advance is to utilize the temporal sequencing of the data in order to extract the cyclic fluxes. To allow for the possibility of fluxes, the distance measure used by LOOPER to reconstruct the manifold is asymmetric. Similar approaches\cite{Talmon2013} have been previously suggested and again it is likely that several specific implementations could effectively achieve this goal.

The chief innovation of LOOPER, however, is conceptual rather than methodologic. In principle, manifolds formed by nonlinear dynamical systems can be arbitrarily complex and high-dimensional. When such a dynamical system is evolved on the basis of its ability to solve a particular task, the repertoire of suitable manifolds and their dimensionality is highly constrained. The main hypothesis motivating the development of LOOPER is that in noisy systems optimized for the performance of a particular task, dynamics can be approximated by a collection of loops. These loops form because noise-driven tangling of trajectories greatly limits information storage. Specifically, the only information that can be reliably stored in a noisy system for an extended period of time is the identity of a bundle of tangled trajectories which we refer to as loop. While each one of the loops can trace an arbitrarily complex path through the phase space, topologically it is a one dimensional object. Thus, the overall collection of loops that together comprise a manifold is a two dimensional object. All that is needed in order to simulate and predict neuronal activation is the identity of the loop and the phase along it. LOOPER can be used to identify these key variables in an unsupervised fashion. Simulations of neuronal activity based on this greatly simplified model are in excellent accord with experimental observations.  

There is significant empirical evidence to support the conjecture that manifolds of evolved dynamical systems can be well approximated by a collection of one dimensional loops. Here, we have demonstrated that dynamics in a working memory task in a primate are well captured by the LOOPER model. The dynamics of neuronal control of a vastly different locomotor behavior in a much simpler nervous system of nematode \textit{C. elegans} was also well approximated by a collection of loops\cite{Kato2015,Brennan2017,Brennan2019a}. Using somewhat different approaches such as jPCA\cite{churchland2012neural} and LFADs\cite{Pandarinath2018}, neuronal activity associated with reaching behaviors in a primate was found to be largely confined to a loop traced out in phase space. Using locally-linear embedding, representation of odors was mapped onto a loop in insects\cite{Stopfer2003}. Dynamics in the head direction circuit of the hippocampus during both sleep and wakefulness was also found to be confined to a loop\cite{chaudhuri2019intrinsic}. 

Loop structures do not appear to be a direct consequence of the biological constraints on the dynamics of the nervous system. Indeed, while activity in the primate brain is clearly different from that in an RNN, the topological description of RNN dynamics trained to perform a working memory task was essentially identical to that in a primate. The common thread appears to be not the specifics of neuronal architecture that gives rise to the dynamics but the demands of the task to which the dynamics are optimized.

There is a compelling theoretical argument (see Supplement for a more rigorous treatment) to suggest that loop structures are likely to be commonly observed in stochastic dynamical systems optimized to perform a task. One argument illustrated in Figure~\ref{Fig:1} is that once trajectories of the system cross, all of the information concerning the history of the system up to that point is erased. Thus, in order to store information, stochastic dynamical systems must organize different sets of trajectories -- each corresponding to a distinct component of the “computation” being performed -- into bundles that remain separate for as long as the information encoded by this trajectory bundle is needed. The system consisting of a single loop, however, cannot respond to changes in the environment. No matter the circumstances, the system will continue to evolve along the same recurrent trajectory. Many biologically salient tasks require changes in behavior on the basis of the stimulus. This ought to lead to branching of the trajectory bundles depending on the stimulus and, critically, the state of the system. Thus, we would expect that when the computation requires a decision point (such as comparing two stimuli) the trajectories would branch out into  distinct bundles for each possible outcome of the comparison. Indeed, this is exactly what we observe in both the primate brain and the recurrent neural network trained to perform a similar task. Thus, the branching pattern can be used to characterize the computation performed by the system.

To further support this assertion we found that, when a slightly modified training set is used, an RNN arrives at a fundamentally different solution to the problem. The RNN gave consistently erroneous answers on some combinations of stimuli but performed adequately on other novel stimulus combinations. We verified that the computation performed by this poorly-conditioned RNN is exactly as predicted by the LOOPER model. Using the branching pattern of the manifold we predict which novel combinations of stimuli will yield consistently incorrect responses. Thus, parallel construction of LOOPER manifolds on biological and artificial networks can reveal both the similarities and differences between different strategies used to solve the same task. The ability to extract the LOOPER manifold, may also endow us with the ability to ask how learning may affect global dynamics of the biological or in silico systems. The LOOPER hypothesis clearly predicts that with training, the initially disparate activity trajectories will coalesce to form trajectory bundles. Understanding how this bundle formation takes place in a dynamical system may inform how learning takes place in a biological system and improve upon training of artificial recurrent neural networks. 

Much of our understanding of the physical world relies on the assumption that a given phenomenon will recur when the circumstances which give rise to it are repeated. If all the circumstances could be simultaneously reproduced, this principle could be universally applied; but this never happens; some of the circumstances will always be missing\cite{poincare1905science}. In an abstract dynamical system, it is not a priori clear which if any of these missing or variable initial conditions are important. In a stochastic dynamical system optimized to perform a particular task, in contrast, it is possible to empirically determine whether differences in the initial conditions along a particular direction in phase space are indeed consequential by examining whether trajectories emanating from two points evolve in a similar fashion. If they do, then the differences in the initial conditions do not carry significant behaviorally salient information and the two trajectories can be safely fused into a single trajectory bundle. LOOPER uses this intuition to extract behaviorally relevant dynamics in an unsupervised fashion solely on the basis of experimental observations.

\begin{methods}
	The method implemented in LOOPER  has three primary steps: asymmetric diffusion mapping, matrix reduction, and loop clustering. The first step involves estimating transition probabilities between points in local neighborhoods. Once a matrix made up of these local transition probabilities is constructed, points with similar transition probabilities are clustered in the second step. This reduced matrix is then clustered again to identify repeating sequences of clusters which form loops in the final model. The overall goodness of fit is then evaluated to validate the clustering algorithm. 
	
	\subsection{Step1: Asymmetric diffusion mapping.}
	
	In everything that follows, we assume that the data are in phase space. The specifics of how phase space can be extracted from experimental observations is described in section \textbf{Latent Space Extraction}. Diffusion mapping is a nonlinear dimensionality reduction technique. Like other nonlinear dimensionality reduction techniques, diffusion maps utilize the fact that nonlinear manifolds are locally Euclidean.  In diffusion maps, distances between points in a local neighborhood are expressed as transition probabilities using a diffusion kernel where the probability of a point, $\mathbf{x}_1$, transitioning to another point, $\mathbf{x}_2$ in one time step is:
	\begin{equation} \label{Eq:P}
	P  (\mathbf{x}_1,\mathbf{x}_2)=exp\left (-\frac{D(\mathbf{x}_1, \mathbf{x}_2)^2}{2 \sigma^2}  \right ),
	\end{equation}
	
	where $D$ is a measure of distance, and $\sigma$ is a normalization term that sets the size of the local neighborhood. 
	\begin{equation} \label{Eq:Plocal}
	P_{local}(\mathbf{x}_1,\mathbf{x}_2) = \left\{\begin{matrix} P(\mathbf{x}_1,\mathbf{x}_2)  & \text{if } \mathbf{x}_2 \in nbh(\mathbf{x}_1) \\ 0 & \text{elsewhere} \end{matrix}\right.
	\end{equation}
	where $nbh(\mathbf{x}_1)$ is the local neighborhood of $\mathbf{x}_1$ and is defined as all points, $\mathbf{x}$ such that $D(\mathbf{x}_1,\mathbf{x}) < 2\sigma$.
	
	The pairwise local transition probabilities between all $n$ points are assembled into a $n \times n$ global transition probability matrix by applying $P_{local}$ to each of the  $n$ experimental observations. This matrix models diffusion with a local neighborhood after one time step. Exponentiation of this transition probability matrix yields diffusion over longer time periods. In this way the global nonlinear structure of the manifold is stitched together from local neighborhoods.  
	
	Before Eq.~\ref{Eq:Plocal} can be applied to the data to assemble the transition probability matrix, distance measure $D$ and an appropriate scaling for the neighborhood $\sigma$ have to be defined. The methodology used to accomplish these goals is described below. 
	
	\subsection{Distance Measure.}
	
	We are specifically interested in modeling dynamical systems. In order to account for the temporal sequencing of the data, we take both the location of the data and its time derivative (velocity) into account when computing distances between points:
	\begin{equation} \label{Eq:Dcombine}
	D_\text{combine} = 1 - (1 - P_{D_{\dot{x}}}) \cdot (1 - P_{D_\mathbf{x}}).
	\end{equation}
	where $P_{D_{\dot{x}}}$ is the normalized difference in velocity, and $P_{D_\mathbf{x}}$ is the normalized difference in position. Differences in position $D_{\mathbf{x}}$ and velocity $D_{\mathbf{\dot{x}}}$ are normalized as follows:
	\begin{equation} \label{Eq:PD}
	P_D = D(\mathbf{x}_1, \mathbf{x}_2) / \max_{\mathbf{x}} D(\mathbf{x}_1, \mathbf{x}),
	\end{equation}
where $D$ is one of the two distance measures as above, and the denominator is the maximum value the distance measure takes over all observed points. For the difference in position we use standard Euclidean distance scaled by the approximation of the local noise (see section \textbf{Estimation of Local Noise}). For difference in velocity (also scaled by the approximation of the local noise) we use the cosine distance: 
\begin{equation} \label{Eq:Dxdot}
D_\mathbf{\dot{x}}  (\mathbf{x}_1,\mathbf{x}_2)=1-\frac{\dot{\mathbf{x}}_1 \cdot  \dot{\mathbf{x}}_2}{\left |  \dot{\mathbf{x}}_1 \right | \left |  \dot{\mathbf{x}}_2 \right |},
\end{equation}
where $\cdot$ is the inner product and $\left | \cdot \right |$ is the magnitude of the vector. Note that $D_{combine}$ is small only when two points $\mathbf{x}_1$ and $\mathbf{x}_2$ are near each other and move in the same direction. Thus, $D_{combine}$ is a natural Newtonian way of defining distances between experimental observations that arise in dynamical systems. 

\subsection{Partitioning the data into local neighborhoods.}

One of the greatest challenges for constructing a diffusion map is selecting the correct size for the local neighborhood in Eq.~\ref{Eq:Plocal}. In this section we describe how this is accomplished in LOOPER. Once the local neighborhood is defined, it has to be rescaled in order to take into account anisotropic diffusion. This rescaling will be addressed in section titled \textbf{Estimation of Local Noise}.

One strategy for defining a local neighborhood is to use $k$-nearest neighbors. An alternative approach sets the size of the point cloud and assigns all the points inside this cloud to a local neighborhood. Choosing appropriate $k$, or  the appropriate neighborhood size, however, is not trivial. This choice is made harder still because there is no guarantee that the optimal $k$, or neighborhood size is the same everywhere. One approach is to choose a different optimal $k$ for each neighborhood. This approach can work but quickly becomes computationally prohibitive. Thus, here we utilized a more computationally efficient strategy that first uses $k$-nearest neighbor approach to estimate local noise and then uses this estimate of noise to define the size of the local neighborhood. 

First, we define a single $k$ for all neighborhoods. This is a free parameter of the method. The primary purpose of choosing $k$ is to select a subset of data which will be used to compute the local noise $\mathbf{\sigma}_{local}$. 

We start with computing the Euclidean distances from the point of interest $\mathbf{x}_t$ to all other data points. The naive approach would be to simply find $k$-nearest points. However, if the experimental observations are sampled finely enough in time, points in the neighborhood of $\mathbf{x}_t$ will most likely be observed around time $\mathit{t}$. We are interested in recurrent trajectories -- dynamics repeatedly observed during execution of the task. To find nearest trajectories, rather than just nearest points, we utilize an approach similar to constructing return maps\cite{teschl2012ordinary}-- a dynamical systems methodology which identifies states to which the dynamical system returns multiple times. Thus, we define a minimum return time, $\tau$ (a parameter in the method). $\tau$ should be sufficiently large to assure that the system first leaves the neighborhood of $\mathbf{x}_t$ prior to returning to it. The first nearest neighbor is the closest point that is separated in time from the point of interest by at least $\tau$. The second nearest neighbor is the next closest point that is separated in time from both the point of interest and the first nearest neighbor by at least $\tau$. This iterative process is repeated until $k$ nearest neighbors have been selected.  The result of this approach is a collection of points that together form a local neighborhood to which the system returns during repeated execution of behaviors. We define this set of points as $\mathcal{N}_{original}(\mathbf{x}_t)$. This subset of points will be now used to estimate local noise $\mathbf{\sigma}_{local}$ around each experimental observation as described in the following section.

\subsection{Estimation of Local Noise.}

Note, the originally symmetric point cloud in Figure~\ref{Fig:1}B is distorted as the system is allowed to evolve in time. The dispersion is preferentially dampened in all directions orthogonal to the flux. To account for this anisotropy, the distances between $\mathbf{x}_t$ and all other points have to be rescaled. In order to take into account the direction of the local flux, we construct an agglomerate neighborhood by combining $\mathcal{N}_{original}(\mathbf{x}_t)$ and all points that are one time step removed (either preceding or following) each point within $\mathcal{N}_{original}(\mathbf{x}_t)$. The result is a 3$k$-sized point cloud which is used to estimate the standard deviation $\mathbf{\sigma}_{local}$. For $N$-dimensional data, $\mathbf{\sigma}_{local} $ is an $N$-dimensional vector $\left\langle \sigma_{local}^{(1)}, \sigma_{local}^{(2)} \dots \sigma_{local}^{(N)} \right\rangle$. This vector is used to rescale distances between the point of interest $\mathbf{x}_t$ and all other points as follows:
\begin{equation} \label{Eq:Dx}
D_{\mathbf{x}}(\mathbf{x}_t, \mathbf{x})=\sqrt {\sum_{i=1}^{N}\left( \frac{{x}_{t}^{(i)}-{x}^{(i)}}{\sigma_{local}^{(i)}}\right )^2},
\end{equation}
where $x^{(i)}$ is the $i$-th dimension of the $N$-dimensional data and $\sigma_{local}^{(i)} $ is the $i$-th dimension of the local noise. This equation corresponds to expressing Euclidean distances between the point of interest $\mathbf{x}_t$ and all other points in terms local $z$-scores. This normalization spherizes local neighborhoods and corrects for the anisotropic elongation of point clouds along the direction of the flux. Because $\mathbf{\sigma}_{local}$ is computed locally for each data point, this normalization allows for changing direction of the flux as it traverses the phase space. We apply conceptually the same normalization to the cosine distance of velocities given in Eq.~\ref{Eq:Dxdot}. The two distance measures used in  Eq~\ref{Eq:Dxdot} and Eqs~\ref{Eq:Dx} have now been defined and can be used to compute $D_{combine}$ (Eq.~\ref{Eq:Dcombine}).

The final step in constructing the local neighborhood is to choose the normalization term, $\sigma$, for Eq.~\ref{Eq:P} which ultimately limits the size of the diffusion kernel. To accomplish this we apply the same $k$-nearest neighbor approach as in section \textbf{Partitioning the data into local neighborhoods} to construct $\mathcal{N}_{scaled}(\mathbf{x}_t)$, using the distance measure $D_{combine}$. We then choose $\sigma$ to be equal to the largest distance between $\mathbf{x}_t$ and elements of $\mathcal{N}_{scaled}(\mathbf{x}_t)$. We now have assembled all the elements necessary to compute the diffusion map (Eq.~\ref{Eq:Plocal}). This map can be used to simulate diffusion after one time step. In the next section, we describe how global structure of the manifold can be reconstructed from local transition probabilities. 

\subsection{Reconstruction of Global Distances.}

In order to increase the robustness of the local neighborhood we symmetrize the diffusion probabilities as follows:
\begin{equation} \label{Eq:Pfinal}
P_\text{final} (\mathbf{x}_1,\mathbf{x}_2) = \text{min} \left[ P_\text{local} (\mathbf{x}_1,\mathbf{x}_2), P_\text{local} (\mathbf{x}_2,\mathbf{x}_1)\right] = P_\text{final} (\mathbf{x}_2,\mathbf{x}_1).
\end{equation}
This allows the global information encoded in pairwise similarities to error check cases in which the local neighborhood estimate of a single point is poorly conditioned.

The diffusion map, $L$, can now be built by taking $P_\text{final}$ for each pair of points in the observed dataset:
\begin{equation} \label{Eq:L}
L_{ij} = P_\text{final} (\mathbf{x}_i,\mathbf{x}_j).
\end{equation}

$L$ only considers local transition probabilities after one time step. In order to explore the global structure of the manifold, we must allow the system to evolve in time and diffuse beyond the local neighborhoods. This is accomplished by first converting $L$ to a Markov matrix:
\begin{equation} \label{Eq:Lmarkov}
L^\text{markov}_{ij} = \frac{L_{ij} }{\sum_j L_{ij}}.
\end{equation}
We can then simulate the time evolution of the system by exponentiating this matrix. The goal of the exponentiation is to repopulate the elements beyond local neighborhoods. Thus, we exponentiate $L^{markov}$ until the number of non-zero elements in a row of $L^{markov}$ is greater than 95\%  (a parameter of the method). Thus, through exponentiation, local neighborhoods are stitched together and global manifold structure of the data is now encoded in: 
\begin{equation} \label{Eq:Lglobal}
L^{global} = (L^{markov})^T,
\end{equation}
where $T$ is the number of exponentiations required to repopulate the matrix.

We now normalize $L^{global}$ to obtain a discrete approximation of the Fokker-Planck operator. Fokker-Planck operator describes the time evolution of a distribution of points driven by a stochastic dynamical system. First, we approximate the steady state distribution, $\pi$, of the system using standard spectral analysis of $L^{global}$. Next, we normalize each element of $L^{global}$ by the steady state distribution,
\begin{equation} \label{Eq:Lfp}
L^{FP}_{ij} = \frac{L^{global}_{ij} }{\sqrt{\pi_i \pi_j}}.
\end{equation}
The resulting matrix is then converted to a Markov matrix, $M$, as before. $M$ is the discrete Fokker-Planck operator of the system (Figure~\ref{Fig:diffusionMap})\cite{nadler2006diffusion}. $M$ quantifies the diffusion of a system after $T$ time steps. 

Unfortunately, standard diffusion maps are not well suited for the extraction of coherent trajectories because they are symmetric by construction. Thus, $M$ will always have the detailed balance condition, and describe purely diffusive processes. We are ultimately  interested in finding coherent trajectories with non-zero flux. The only flux that satisfies the stationarity assumption is divergence free and cyclic\cite{Wang2008}. Such fluxes correspond to rotations in state space. These rotations occur when the transition probability matrix has complex eigenvalues. Conceptually, this suggests a modification of the standard diffusion kernel by centering the kernel not on a point, $\mathbf{x}_t$, but on its next observed point, $\mathbf{x}_{t+1}$,
\begin{equation} \label{Eq:P2}
P  (\mathbf{x}_t,\mathbf{x})=exp\left (-\frac{D(\mathbf{x}_{t+1}, \mathbf{x})}{2 \sigma^2}  \right ).
\end{equation}
To achieve this, we remove the first column and last row of $M$. This is akin to shifting the transition probabilities to the upper one-off diagonal, and has the effect of making the most likely transition from the current time point, $\mathbf{x}_t$, always be the next observed time point, $\mathbf{x}_{t+1}$. The resulting asymmetric matrix is a discrete approximation of a stochastic dynamical system that contains both diffusion and flux. By reducing this matrix we will ultimately be able to approximate the dynamics of the system. The next step in the LOOPER algorithm involves simplifying such matrices. The methodology that accomplishes this is described in the next section. 

\subsection{Step 2: Reducing the matrix}

We begin by renormalizing the diffusion map so that similar points are grouped together. Thus, the overall goal of the next step is to convert the $n \times n$ matrix $M$ spanned by experimental observations to a smaller $C \times C$ matrix where each element corresponds to a cluster of similar points.  

The goal of the matrix renormalization is to condense the asymmetric diffusion map into a minimal number of states without destroying the information contained in the original matrix. The optimal number of states can be found using the minimum description length model\cite{beretta2018stochastic}, which balances the number of parameters of the model and the information loss of the model:
\begin{equation} \label{Eq:MDL}
\text{MDL} = I_\text{loss} + \frac{k}{2}\log{\frac{n}{2\pi}},
\end{equation}
where $I_\text{loss}$ is how much worse the new model $M_{reduced}$ performs compared to the original model $M$, $k$ is the size of $M_{reduced}$  and $n$ is the size of $M$. $M_{reduced}$ is defined as follows:
\begin{equation} \label{Eq:Mreduced}
M_{reduced}(i,j) = \sum_{a \in c_i, b \in c_j} \frac{M(a,b)}{| c_i |}.
\end{equation}
Where $a, b$ are indices of data points in $M$, $c_i$ is the $i$-th cluster of points in $M_{reduced}$, and $\left | \cdot \right |$ is the number of elements in the cluster. Thus, the reduced matrix has $k = C^2$ elements, where $C$ is the number of clusters in $M_{reduced}$. Further, every datapoint indexed by $a$ of $M$ maps to a cluster $c_a$ in $M_{reduced}$ by $a \mapsto c_a$. To determine how much information is lost by such clustering, we reconstruct the original matrix from its clustered version.
\begin{equation} \label{Eq:Mreconstruct}
M_{reconstruct}(a,b) = M_{reduced}(c_a, c_b). 
\end{equation}
The information lost in this $M_{reconstruct}$ is used to optimize the number of clusters in $M_{reduced}$.

We wish to enforce low information loss over all observed states. For computational efficiency, we constrain the upper bound of information loss and  consider the states with the worst information loss. Thus, we approximate the information loss due to clustering of data points as:
\begin{equation} \label{Eq:Iloss}
I_\text{loss} \approx n \cdot \max_{t} Q_\text{KL} (0.95),
\end{equation}
where $Q_\text{KL}$ is the 95\% quantile of the following value over all observed starting states, $i$:
\begin{equation} \label{Eq:Dkl}
D_\text{KL}\left (M(t | t_0 = i) \| M_{reconstruct} (t | t_0 = i) \right).
\end{equation}
Here $D_\text{KL}$ is the Kullback-Leibler divergence (a measure of how much information about the distribution is lost by applying a given approximation), $M(t | t_0 = i)$ is the time evolution of the distribution over all observed states when starting from state $i$, and $M_{reconstruct}(t | t_0 = i)$ is the time evolution of the distribution over observed states of the reconstructed matrix when starting from state $i$. The time evolution of a Markov matrix starting from state $i$ is given by:
\begin{equation} \label{Eq:Pt}
P(t|t_0=i) = P^t \cdot e_i
\end{equation}
Where $e_i$ is a vector with a 1 for the $i$-th element and zeros everywhere else. Note that $i$ refers to the observed state, and not the corresponding reduced cluster. We maximize $I_{loss}$ over time in order to account for the possibility that the reduced approximation can have initially low information loss that grows over time. This can occur when geometrically similar but topologically distinct clusters  are fused. In order to keep the computation tractable, we test $t$ from 1 to 5 (a parameter of the method). When then pick the number of clusters such that $MDL$ is minimized over a range of putative cluster counts (a parameter in the method). 

The only parameter that needs to be defined is the similarity between states within the original matrix $M$. This similarity is defined as follows:
\begin{equation} \label{Eq:S}
S_{ij} = 1 - \text{corr}(M_i, M_j),
\end{equation}
where $M_i$ is the $i$-th row of the original diffusion map $M$. Intuitively this means that if the the distribution of states observed in the next time step starting from state $i$ and $j$ is similar, then states $i$ and $j$ are similar. 
Hierarchical clustering based on $S$ was implemented using  the MATLAB linkage command.  The end result of this procedure is a new $C \times C$ transition probability matrix that instead of experimental observations contains cluster ids as its entries (Figure~\ref{Fig:clusteredMap}). This renormalization tremendously simplifies discovery of loops described in the next section.    

\subsection{Step 3: Loop Finding}

The objective in this step is to further simplify the $C \times C$ matrix $M_{reduced}$ by describing the dynamics that arise in it as a collection of interconnecting  loops. The dynamics given by $M_{reduced}$ is a sequence of cluster ids. A loop, therefore is a sequence of cluster ids that begins and ends in the same cluster. We can conceptualize such a loop as a word, $W$, where each letter corresponds to a cluster id. Therefore, a natural way to compare different $W$ is to use an edit distance.  We use this general approach in LOOPER. We first identify $W$ given by the $M_{reduced}$ . We then apply an edit distance to combine such $W$ into loop clusters $C_W$. The clustering procedure is then optimized. The specific procedure for accomplishing this clustering and optimization is outlined below. 

\subsection{Finding Loops}

We start with $M_{reduced}$- a $C \times C$ diffusion map where each entry corresponds to a cluster, $c$ formed by the experimental observations. Using this information, we construct a bandwidth matrix $B$ where each entry is the number of times that a data point observed to be in $c_i$ at time $t$ is found in $c_j$ at time $t+1$ :  
\begin{equation} \label{Eq:B}
B_{i,j\neq i} = \left | \{\mathbf{x}_t \in c_i,\mathbf{x}_{t+1} \in c_j\} \right |,
\end{equation}
where $\left | \cdot \right |$ is the number of elements in the set and $\mathbf{x}$ is an experimental observation. 

We next find the shortest path starting and ending in $c_i$ -- the shortest loop-- using standard Dijskstra shortest path algorithm\cite{dijkstra1959note} applied to $B^{-1}$. Thus, we find $C$ shortest loops $W$, one for each cluster $c$. Clearly, some of these loops will overlap. For example, if a sequence of clusters $\{c_1, \dots, c_n \}$ is a loop, then any shift permutations of this sequence is also a loop. In order to quantify similarity between two loops, we therefore employ an edit distance. However, standard application of an edit distance cannot take into account the fact that some clusters are more similar to each other than others. Thus, prior to the application of an edit distance, similarity between clusters needs to be defined. This is accomplished as follows:
\begin{equation} \label{Eq:Scluster}
S_\text{cluster}(i,j) = \frac{M_{\text{cluster}_i} \cdot M_{\text{cluster}_j}}{\left \| M_{\text{cluster}_i} \right \| \left \| M_{\text{cluster}_j} \right \|}
\end{equation}
where $\left \| \cdot \right \|$ is the Euclidean norm, and $M_{\text{cluster}_i} $ is the $C \times n$ matrix of mean transition probabilities computed  across all points that belong to a given cluster $c_i$ to all $n$ experimental observations:
\begin{equation} \label{Eq:Mcluster}
M_{\text{cluster}_i} = \frac{1}{|c_i|}\sum_{t \in \text{c}_i }M_t,
\end{equation}
where $| \cdot |$ is the number of elements, and $M_t$ is the $t$th row of $M$. $S_{cluster}(i,j)$ is analogous to $S$ (Eq.~\ref{Eq:S}) in that it quantifies the similarity in transition probabilities of all states that make up cluster $i$ and cluster $j$. Now, equipped with the cluster similarity measure, we can readily compare loops, $W$ formed by sequences of clusters as follows:
\begin{equation} \label{Eq:Sloop}
S_\text{loop}(i,j) = \prod_{c_k \in W_i} \max_{c_l \in W_j} S_\text{cluster}(c_k, c_l)
\end{equation}
Note that when the same cluster $c$ is found in both loops $i$ and $j$ $S_{loop}(i,j)$ be will always be 1. Therefore $S_{loop}$ readily identifies shift permutations. When clusters assigned to $W$ are distinct, $S_{loop}$ is heavily influenced by the similarities between clusters, $S_{cluster}$. 

We use $S_{loop}$ to cluster the system into a set of loops. The confidence in the estimate of  $S_{loop}$ depends strongly on how many times a particular sequence of states is experimentally observed. Thus, we bias the clustering to merge the least traversed loops first. This is accomplished by dividing each entry in $S_{loop}$ by the loop bandwidth:
\begin{equation} \label{Eq:Bloop}
B_{loop}(i) = \sum_{c_j \in W_i} \frac{B(c_j, c_{j+1})}{|W_i|} . 
\end{equation}
Finally, we symmetrize the similarity matrix by:
\begin{equation} \label{Eq:Sloopfinal}
S_\text{loop}(i,j) = \text{max}(S_\text{loop}(i,j), S_\text{loop}(j,i)). 
\end{equation}
This matrix is then submitted to hierarchical clustering. The output of this clustering is a set of loop clusters $C_W$. Each $C_W$ is a collection of ordered sequences $W=\left \{ c_i \dots c_n, c_i \right \} $ that begin and end in the same cluster of points $c$. The optimization of this clustering procedure and the assembly of the loop clusters into the final model is described in the next section.

\subsection{Assembling Loops into the complete model.}

To optimize the number of loop clusters, $C_W$  we use the following heuristic. We would like to choose the number of loop clusters such that the overall model faithfully recapitulates the experimental data. This validation requires two steps. First, an agglomerate average loop has to be constructed from all the loops that together comprise a loop cluster $C_W$. This operation is done solely on the basis of loops $W$. In the second step, each phase along $W$ has to be mapped back from the abstract manifold space onto experimental observations. This last step uses the coordinates of the experimental observations $\mathbf{x}$ that make up each cluster $c$ within $W$.  The number of loop clusters is optimized over a pre-specified range (a parameter of the method) such that this two step process faithfully reconstructs the data. The specifics of how this optimization is accomplished is outlined below.

For each putative number of loop clusters we first need to approximate each $C_W$ by an average loop. 
This is not trivial, as different loops $W$ are typically shift permutations of each other. Thus, in order to appropriately average them, the phases of all $W$ within a loop cluster need to be aligned first. 

We arbitrarily define 0 to be the most commonly observed cluster, $c_{common}$ amongst all the loops, $W \in C_W$. Now, we define the start for each of each loop as cluster $c_{start}$ such that $S_{cluster}(c_{start}, c_{common})$ is minimized. Because, by definition, each $W$is a loop, given the start state $c_{start}$ the rest of the point clusters $c$  that comprise each $W$ can be assigned a phase, $\theta_{cluster}$ between 0 and $2\pi$ using linear interpolation. The phase  resolution $\Delta\theta_{cluster}$ therefore is given by the number of states that comprise each $W$.     

What we would like to ultimately achieve is a more continuous definition of phase. To accomplish this, we average locations of data points with respect to the phase of the loop. This is performed using a standard approach consisting of first defining a phase bin with desired granularity and averaging the position of all data points that fall into the phase bin. The procedure for performing this mapping is outlined below.

We first pick the number of phase bins (a parameter of the method) and construct a Gaussian window centered at each bin. To define the width of this Gaussian window we compute a weighted average of the number of clusters that comprise each loop in $C_W$. We weigh loop $W$ in $C_W$ in a manner similar to Eq.~\ref{Eq:Bloop}.   
\begin{equation} \label{Eq:OmegaW}
\Omega_{W}(i) = \sum_{c_j \in W_i} M_{reduced}(c_j, c_{j+1})
\end{equation}
We can now define the width of the Gaussian phase window as follows:
\begin{equation} \label{Eq:simgaphase}
\sigma_{phase}=\frac{\pi}{\sum_{W_i \in C_W}\Omega_W(i) \cdot |W_i|}
\end{equation}
where $|\cdot|$ is the number of clusters, $c$  that comprise the $i$-th loop $W$. Thus, $\sigma_{phase}$ is chosen to be approximately $1/2$ the average $\Delta\theta_{cluster}$. 

Now that we have defined a phase bin, we can interpolate the coarse description of phase  given by $\theta_{cluster}$ and identify positions in the data space that are associated with each phase bin. 
\begin{equation} \label{Eq:barxtheta}
\bar{\mathbf{x}}_{\theta(i)} = \sum_{j \in C_W} exp\left (\frac{-(\theta(i) - \theta_{cluster}(j))^2}{2 \sigma_{phase}^2}\right ) \cdot \Omega_{cluster} (j) \cdot \bar{\mathbf{x}}_{j}
\end{equation}
where $\bar{\mathbf{x}}_j$ is the average position of all points in cluster $c_j$, $\theta(i)$ is the $i$-th phase bin and, finally $\Omega_{cluster}$ is the weight of the cluster defined as follows: 
\begin{equation} \label{Eq:Omegacluster}
\Omega_{cluster}(j) = \sqrt{\left | c_j \right |} \cdot \sum_{\{W_k | W_k \in C_W, c_j \in W_k\}} \Omega_W(k)
\end{equation}
This equation is scaling the contribution of each point cluster $c$ by its size and by how often it appears in different loops $W \in C_W$. 

The ultimate result of the above is that we approximated each $C_W$ by an average loop. Furthermore, the position along this loop is defined in terms of finely grained phase $\theta$. Finally, each $\theta$ is now associated with the average location of the experimental data $\bar{\mathbf{x}}_{\theta(i)}$. Thus, each average loop $\overline{W}$ is now defined as an ordered sequence $\left \{\bar{\mathbf{x}}_{\theta(1)} \dots \bar{\mathbf{x}}_{\theta(N)} \right \}$ where $N$ is the number of phase bins.  

This weighted approximation tends to underestimate the curvature of the loops (Figure~\ref{Fig:loopClusters}). Thus, in order to construct a more accurate model we refine the average loop, $\overline{W}$. First, to make calculation of distances in high dimensional spaces more robust, we perform principal component analysis (PCA) on  $\overline{W}=\left \{\bar{\mathbf{x}}_{\theta(1)} \dots \bar{\mathbf{x}}_{\theta(N)} \right \}$. Both $\left \{\bar{\mathbf{x}}_{\theta(1)} \dots \bar{\mathbf{x}}_{\theta(N)} \right \}$ and $C_W$ are then projected onto the first three PCs. Euclidean distance between each point in $C_W$ and $\left \{\bar{\mathbf{x}}_{\theta(1)} \dots \bar{\mathbf{x}}_{\theta(N)} \right \}$ is now calculated. The closest $\bar{\mathbf{x}}_{\theta(i)}$ is found for each point in $C_W$. Finally, we adjust the location of $\bar{\mathbf{x}}_{\theta(i)}$ to be the mean position of all points that have this $\bar{\mathbf{x}}_{\theta(i)}$ as their closest bin. This produces an average loop that is able to more closely track the shape of the loops.

The result is that each point is assigned a position in model space $\mathcal{M}=\left \{\theta(i),C_W \right \}$. The final step used to simulate model dynamics is to assemble a transition probability matrix $M_{model}$ where each element $(i,j)$ is given by:
\begin{equation} \label{Eq:Mmodel}
M_{model}(i,j)=P(\mathcal{M}_t = i | \mathcal{M}_{t+1}=j )
\end{equation}
where the conditional probability $P$ is empirically derived from the data and $i$ and $j$ correspond to  unique state in model space defined by $\left \{\theta,C_W \right \}$. This transition probability matrix, along with the mapping from state $\{\theta, C_W\} \mapsto \bar{\mathbf{x}}_{i}$, makes up the minimal model of dynamics (Figure~\ref{Fig:1}F). This model emits a sequence of $\bar{\mathbf{x}}_{1}, \bar{\mathbf{x}}_{2} \dots \bar{\mathbf{x}}_{N}$ which can be directly compared to the sequence of experimental observations $\mathbf{x}_1, \mathbf{x}_2 \dots \mathbf{x}_N$. 

The goodness of fit is computed by approximation of the log likelihood of the model. For each point in the observed data we compute a measure of how likely it is to observe a specific transition from $\mathbf{x}_t$ to $\mathbf{x}_{t+1}$:
\begin{equation} \label{Eq:scoret}
score_t = \min_{i \in \mathcal{M}, j \in \mathcal{M}} log(D(\mathbf{x}_t, \bar{\mathbf{x}}_i) \cdot D(\mathbf{x}_{t+1}, \bar{\mathbf{x}}_j)) / M_{model}(i, j).
\end{equation}
Where $D(\mathbf{x}_t,\bar{\mathbf{x}}_i)$ is the Euclidean distance from the observed point $\mathbf{x}_t$ to $\bar{\mathbf{x}}_{i}$.  We minimize the $score_t$ over all possible combinations of $i$ and $j$ in order to find the best possible combination of states and transition probabilities in $M_{model}$ to describe the experimentally observed data's transition from $\mathbf{x}_t$ to $\mathbf{x}_{t+1}$. 

Thus, better scores occur when both $\mathbf{x}_t$ and $\mathbf{x}_{t+1}$ are close to some phase bin $\bar{\mathbf{x}}_{i}$ of the model, and the transition between those two states, $M_{model}(t,t+1)$, is highly probable. The total validation score is the mean score over the entire time series of experimental observations.

We optimize the number of loops by fixing the number of states in $\mathcal{M}$  (a parameter of the method).  We iterate this validation procedure for each putative number of clusters, $C_W$ and choose the final number of clusters such that the total validation score is minimized (Figure~\ref{Fig:loopClusters}). This concludes the process that is used by LOOPER in order to deconstruct the dynamics into a collection of interconnected loops. 

\subsection{Latent space extraction}

Construction of the LOOPER manifold relies heavily on uncovering the appropriate phase space of the data. When working with a complete set of latent variables (phase space) there is a one to one mapping between the position of the system, $\mathbf{x}$ and the flux of the system $\mathbf{F}(\mathbf{x})$. Commonly, the latent variables that make up the phase space are missing or hidden. These missing latent variables can cause a degeneracy in the equations of motion of the system where the same $\mathbf{x}$ gives different values of $\mathbf{F}(\mathbf{x})$. When this occurs, information about the coherent trajectories in the system are lost and LOOPER will fail to find the corresponding loops -- it may instead find a stochastically dominated version of the loops merged together. Thus, an important preprocessing step is to make sure that the data contains all the information required to find the latent variables.

There are many techniques for latent variable extraction such as LFADS\cite{Pandarinath2018}, Hidden Markov Models\cite{Vilares2011}, MIND\cite{Low2018} and NNMF\cite{Anderson2014}, etc, and any can be used in preprocessing before applying LOOPER to the data. LOOPER itself has delay embedding built in. Delay embedding is a method for latent space extraction that relies on increasing the number of independent measurements of the data in order to extract a topologically equivalent representation of the underlying dynamics. To extract these independent measurements, a delay time $\tau$ is chosen such that the autocorrelation between point $\mathbf{x}_t$ and $\mathbf{x}_{t+\tau}$ is small -- implying independent measurements. Several of these time delayed observations are added to each observed time point in the data. According to the Whitney embedding theorem the number of independent measures required to fully reconstruct the underlying dynamics is $2n-1$, where $n$ is the number of dimensions of the manifold. Note that the Whitney embedding theorem assumes that there is no information overlap in the delayed measurements, so in practice typically more than the minimum number of delays are required. There are many possible techniques for choosing the ideal number of delays and the correct value of $\tau$. LOOPER leaves these values as free parameters. Generally, one can over embed and rely on LOOPER's built in dimensionality reduction to remove excess dimensions. Takens' theorem guarantees that the delay embedded dynamics will be topologically equivalent to the underlying dynamics, thus implying that the loop structure will not change.

\subsection{Preprocessing of primate data}

Preprocessing of the primate data is done using the same pipeline as in Kobak  et al\cite{Kobak2016}. Spike rasters are converted to firing frequencies by using a 30ms Gaussian sliding window. Only trials with average firing frequencies between 5Hz and 50Hz are considered. For the data in Figure~\ref{Fig:2} we restrict the analysis to  neurons observed on at least 10 trials in each condition (179 neurons over 6 trial conditions total).

Neurons are recorded in separate sessions. Thus, we bootstrap over trials in order to create randomized pseudotrials which consist of time synchronized firing of all neurons using the stimulus times (Figure~\ref{Fig:2}B). Each pseudotrial is constructed by taking the mean firing of each neuron during a given condition averaged over $n/2$ bootstraps randomized with replacement. Here, $n$ is the number of trials for that particular neuron and condition. For Figure~\ref{Fig:4}B the model is trained on trials with odd trial indices, and validated on trials with even indices (Figure ~\ref{Fig:validation}).

\subsection{Validation of topological structure}

The validation data is projected onto the LOOPER manifold by finding the best model state ID for each observed time point. For each time point, the locally z-scored distance to each cluster is found:
\begin{equation} \label{Eq:Dlocal}
D_{local}(t,i) = \left \| (\mathbf{x}_t - \bar{\mathbf{x}}_i) / \mathbf{\sigma}_i \right \|^2_2.
\end{equation}
Where $t$ is the time of the observed data point, $i$ is the index of $\mathcal{M}$, $\mathbf{x}_t$ is an experimental observation at time  $t$, $\bar{\mathbf{x}}_i$ and $ \mathbf{\sigma}_i$ are  the mean  and standard deviation  of all points in  $\mathcal{M}_i$ respectively, and $\left \| \cdot \right \|^2_2$ is Euclidean distance. To assign $\mathbf{x}_t$ to $\bar{\mathbf{x}}_i$, the minimum of  $D_{local}$ over $\mathcal{M}$ is found. Each time point can then be unequivocally assigned to the phase and the loop identity $\bar{\mathbf{x}}_i \mapsto \left\{\theta, C_W\right\}_i$. Once each time point is assigned a loop ID the topological structure can be plotted as in Figure~\ref{Fig:validation}.

The validation accuracies reported in the main text were computed as the percent of validation data points that mapped onto the same loop ID as the original LOOPER manifold over a specific time window. The time windows were chosen to align with their computational counterparts. Classifying F1 $\rightarrow$ mean accuracy between 1.39s and 3.19s, remembering F1 $\rightarrow$ minimum accuracy between 1.39s and 3.19s and comparing F2 to F1 $\rightarrow$ mean accuracy between 3.89s and 5.89s. Equivalent time frames were taken for the RNN. The p-values were calculated by comparing the validation accuracy to a null model produced by randomly assigning loop IDs from a uniform distribution.

\subsection{Training of Recurrent Neural Networks}
The recurrent neural networks in Figure~\ref{Fig:4}C and Figure~\ref{Fig:5} are trained using PyTorch using Adam with 10000 randomized samples. Inputs are a time series of scalar values (0, F1 or F2) corrupted by Gaussian noise with a sigma of 1.5. The loss function is standard cross entropy on the 3 output classes ("No output", "Greater than", and "Less than") weighted so that the target outputs ("Greater than" or "Less than") contribute the same amount as the "No output" cases. This means that the network must match the given target of "No output" for all time points other than the short target window directly after F2 (Figure~\ref{Fig:3}A). During this  window it is heavily penalized for missing the target output. The models are trained until they have a loss of less than 0.01. Classification accuracy of the network is assessed by taking the mode of the outputs during the target window directly following F2. 

The networks consists of 100 fully connected LSTM units. In order to increase the biological realism of the network we inject Gaussian noise with a sigma of 0.1 into both the hidden and cell state of the LSTM units at each time step. The values of the hidden and cell states have a standard deviation of 2.3 over all observations, so the noise is roughly 4\% of the signal.

\subsection{Simulating LOOPER model with inputs}
In order to construct a complete LOOPER model of a recurrent neural network we augment the unit activity of the RNN with both its inputs and outputs. This results in a $n+2 \times t$ matrix where $n$ is the number of hidden variables, and $t$ the number of timesteps. Without inputs,  simulation of the LOOPER model is exactly like simulation of a Markov matrix. At each time step find the current state of the simulation and transition to a new state with probability equal to the corresponding row of the matrix. With the addition of inputs, we must consider not only the current state of the system, but also the current input to the system. This is done by weighting the transition probabilities of the current state by the input value:
\begin{equation} \label{Eq:wi}
w_i = exp\left(\frac{-(I(t) - \bar{i_i})^2}{2\sigma_i^2}\right),
\end{equation}
where $i$ is the next state ID, $I(t)$ is the current input, $\bar{i_i}$ is the mean input value of state $i$ and $\sigma_i$ is the standard deviation of the input at state $i$. This value is clipped to 0 if less than $exp(-2)$. However, due to the precise temporal timings of inputs in the task, inputs are sometimes missed. This can occur when the simulated system traverses too quickly and randomly jumps out of a state that should only be transitioned out of during inputs. In order to help alleviate this issue we allow some leeway in the timing of inputs.

If the inputs into the system are such that no valid transitions exist from the current state (i.e. all transition probabilities are less than $exp(-2)$), then we draw from the previous state of the simulation instead. This can repeat several times so that the simulation can look several time steps into the past for the most recent state in which a valid input exists. For the simulations in Figure~\ref{Fig:5}C backtracking occurs on only 5\% of time steps and the mean backtracking time is 3.7 time steps (max 12). Thus, this method still produces a good approximation of the true time evolution of the system

\subsection{Extending LOOPER to work with trial data}
LOOPER's loop clustering algorithm specifically assumes that the data forms complete loops in phase space. In the case of short trial-based recordings this may not be the case. In order to recuse this assumption we manually add a single hidden state to the LOOPER model. This hidden state represents all configurations of the system that are not observed in the data. Thus, this state cannot be transitioned to in the observed data. Instead, we assume that every trial transitions to this state after the last observed time point, and that the first time point of a trial transitions from this hidden state. In this way, each trial can be considered a closed loop beginning and ending in the hidden state.

\end{methods}




\begin{addendum}
	\item We thank Rui Pei, Adeeti Aggarwal, Guillermo Cecchi, Marcelo Magnasco, Drew Hudson, Tom Joseph, and Max Kelz for critically reading the manuscript.
	\item[Competing Interests] The authors declare that they have no
	competing financial interests.
	\item[Correspondence] Correspondence and requests for materials
	should be addressed to A. Proekt~(email: Alexander.Proekt@pennmedicine.upenn.edu).
\end{addendum}


\begin{table}[]
	\caption {Parameters used in LOOPER models} \label{Tab:parameters} 
	\begin{tabular}{lllllllllllllll}
		& \multicolumn{4}{c}{Preprocessing}                                                                                                & \multicolumn{4}{c}{Diffusion mapping}                          & \multicolumn{2}{c}{Model reduction}              & \multicolumn{2}{c}{Find loops}        \\
		System & \rotatebox[origin=l]{90}{Use z-score} & \rotatebox[origin=l]{90}{Smoothing} &\rotatebox[origin=l]{90}{Embed delays} & \rotatebox[origin=l]{90}{Embed count} & \rotatebox[origin=l]{90}{Neighbor count} & \rotatebox[origin=l]{90}{Use local dims.} & \rotatebox[origin=l]{90}{Repop. density} & \rotatebox[origin=l]{90}{Min. return time} & \rotatebox[origin=l]{90}{Distance measure} & \rotatebox[origin=l]{90}{Max. check time} & \rotatebox[origin=l]{90}{Total state count} & \rotatebox[origin=l]{90}{Use terminal state} \\
		Fig. 1 Example & F                           & 1                             & 1                                & 1                               & 10       & T               & 0.95           & 10               & corr.            & 5                 & 100             & F            \\
		Primate    & F                           & 2                             & 2                                & 11                              & 7        & T               & 0.95           & 10               & corr.            & 5                 & 200             & T            \\
		RNN (Good) & F                           & 1                             & 2                                & 10                              & 7        & T               & 0.95           & 10               & corr.            & 5                 & 300             & T            \\
		RNN (Poor) & F                           & 2                             & 2                                & 5                               & 7        & T               & 0.95           & 10               & corr.            & 5                 & 300             & T            \\
		Lorenz     & F                           & 2                             & 0                                & 0                               & 20       & T               & 0.95              & 0               & corr.            & 5                 & 100             & F            \\
		
	\end{tabular}
	\caption*{
		For cluster and loop count parameters (not listed in table) we use a range of values such that a minimum BIC value is observed in the middle of the range. Boolean values are listed as T for true and F for false.
	}
\end{table}

\renewcommand{\thefigure}{S\arabic{figure}}
\renewcommand{\theequation}{S\arabic{equation}}
\graphicspath{{SupplementFigures/}}

\clearpage

\section*{Supplement}

\subsection{Mathematical motivation for topological modeling of adaptive dynamical system using loops.}

The underlying biophysics of most biological processes are inherently stochastic in nature. Nonetheless, neuronal computations relying on these stochastic processes can be remarkably consistent. How do such systems reconcile the stochastic nature of their biophysics with robustness of their responses? We show when a stochastic dynamical system is adapted to perform a specific task dynamics are surprisingly limited. This allows for a parsimonious description of the dynamics of the system that is at once quantitatively accurate and qualitatively interpretable.

Coherent trajectory bundles which we call loops arise when the stochastic dynamical system given by $\dot{\mathbf{x}}=\mathbf{f}(\mathbf{x})+\epsilon$ actively stabilizes the noise along all directions orthogonal to the local direction of time evolution of the system given by $\mathbf{f(x)}$. We can study the stability of the system by linearization around each location in state space:
\begin{equation} \label{Eq:xdot}
\dot{\mathbf{x}} \approx \mathbf{f}(\mathbf{x}_\circ) + D_\mathbf{f}(\mathbf{x}_\circ) \cdot (\mathbf{x} - \mathbf{x}_\circ),
\end{equation} 
where $\mathbf{x}$ is a point in the local neighborhood of $\mathbf{x}_\circ$, $\mathbf{f}(\mathbf{x}_\circ)$ is $\mathbf{f}$ evaluated at $\mathbf{x}_\circ$, and $D_\mathbf{f}(\mathbf{x}_\circ)$ is the Jacobian matrix of $\mathbf{f}$ evaluated at $\mathbf{f}(\mathbf{x}_\circ)$. We can now define the divergence of points in the local neighborhood $\mathbf{x}_\circ$ as $\Delta \mathbf{x} \equiv (\mathbf{x} - \mathbf{x}_{\circ})$. The time derivative of the divergence is then:
\begin{equation} \label{Eq:dotdelta}
\dot{\Delta \mathbf{x}} \approx D_\mathbf{f}(\mathbf{x}_{\circ}) \cdot \Delta\mathbf{x}.
\end{equation} 
The solution to this differential equation can be written in terms of eigenbasis of $D$,
\begin{equation} \label{Eq:deltat}
\dot{\Delta \mathbf{x}(t)} \approx \sum_i c_i e^{\lambda_i t} \mathbf{\nu}_i,
\end{equation} 
where $\lambda_i$ and $\mathbf{\nu}$ are the $i$-th eigenvalue and eigenvector of the Jacobian, respectively. $c_i$ is the projection of the starting displacement, $\Delta \mathbf{x}_0$, onto the $i$-th eigenvector. If $\lambda_i$ has a positive real part then $\Delta \mathbf{x}$ grows over time, and any noise along the associated eigenvector is amplified in time. Conversely, if the eigenvalue has a negative real part then noise along the associated eigenvector is damped out. Damping will cause noisy trajectories starting from nearby locations in phase space to coalesce. Thus, if real parts of all eigenvalues of Eq.~\ref{Eq:deltat} are negative then all noise will dampen over time. If this holds for all points in the local neighborhood then all trajectories starting in the neighborhood of $\mathbf{x}_\circ$ will follow a similar path dictated by $\mathbf{f}(\mathbf{x})$ (Eq.~\ref{Eq:xdot}). Thus, all points starting out in the neighborhood of $\mathbf{x}_\circ$ will follow a similar path thus giving rise to a coherent trajectory from $\mathbf{x}_\circ$ to $\mathbf{x}_1$. If the neighborhood around $\mathbf{x}_1$ is similarly stable, then the coherent trajectory bundle proceeds to $\mathbf{x}_2$, etc. In this way locally stable neighborhoods connected by the flux $\mathbf{f}(\mathbf{x})$ can be daisy chained together to give rise to a coherent trajectory which will be observed as consistent time evolution of the system even in the presence of noise. Thus, formation of trajectory bundles relies critically on stability of a local neighborhood.

If $\mathbf{f}(\mathbf{x}_\circ) = 0$ then the system is at a fixed point. If the system is stable (as above) around $\mathbf{x}_\circ$, then noise will be quickly damped out and the system will remain in the fixed point until acted upon by an external stimuli, or until sufficient noise ejects the system from the stable local neighborhood. A system consisting solely of fixed points will behave in a purely stochastic fashion. For instance a Hopfield network is an example of a dynamical system that  tends to a fixed point where the dynamics of the system are stabilized. Stabilization of the fixed point can be used to store memories and use this memory to modulate its reaction to future stimuli. However, once the system arrives at the fixed point, the memory of the trajectory leading up to it is erased. Thus, such systems are not readily able to respond to stimuli by performing complex sequences of behaviors with fixed phase relationship. Fixed points are just the simplest example of a coherent trajectory that consists of a single stable neighborhood. If $\mathbf{f}(\mathbf{x}_\circ)$ is non-zero, the system will move away from one local neighborhood towards the next neighborhood $\mathbf{x}_1 \dots \mathbf{x}_n$ in the direction specified by $\mathbf{f(x)}$ at each one of the local neighborhoods. Because states of the system are mapped onto behavioral space, evolution along the coherent trajectory bundle will give rise to a consistent sequence of behaviors that originate from state $\mathbf{x}_\circ$. Thus, upon arrival to $\mathbf{x}_\circ$, the system will begin to execute a consistent sequence of behaviors characterized by constant phase relationships. Thus, it appears that having non-fixed point trajectories may be advantageous for biological systems.  

A critical question that needs to be answered in order to develop analytic techniques for analyzing brain dynamics is the dimensionality of the trajectory bundles likely to exist in biological systems. Here we argue that these bundles are limited to 0-dimensional (fixed point) and 1-dimensional (trajectory) objects. To see this, envision a dynamical system with a 2-dimensional limit cycle such as a torus. By definition of a stable limit cycle the directions tangent to the surface must not be stabilizing (in fact, the largest real part of the Jacobian matrix’s eigenvalues will always be 0). In general there will be $n$ unstable directions tangent to the hypersurface, where $n$ is the dimensionality of the limit cycle. The flux will also be tangent to the limit cycle, which means that the orthogonal subspace will contain $n$-1 dimensions that are tangential to the limit cycle, and thus $\lambda_{max} = 0$ as long as $n > 1$ (Figure~\ref{Fig:noise}). Therefore, when modeling the responsive components of stochastic dynamical systems we need only consider 1-dimensional trajectories, as any higher-dimensional structures can be well approximated by purely stochastic processes along the remaining, unstable, dimensions. 

What remains is to explore the constraints on the global structure of these coherent trajectories, and how they might fit together. Returning to the equations of motion (Eq.~\ref{Eq:1}) we note that due to the noise term it is difficult to draw conclusions about the system from the trajectory of a single particle. We will instead treat the system as a whole by studying the evolution of the probability distribution of states of the system observed at time, $P(\mathbf{x},t)$. 

We assume that the system is stationary. Specifically, this means that the equations of motion (Eq.~\ref{Eq:1}) are constant during the  observed time period. This is not strictly true for most biological systems as the system will adapt to its environment, but can be invoked in a piecewise fashion if the observation period is short relative to the adaptation rate. Mathematically, there are a few conditions that must be met in order for a steady state solution to exist. The probability distribution must satisfy the natural boundary condition, $\lim_{\left | x \right | \rightarrow \infty } P^n(\mathbf{x}) = 0$, where $P^n$ represents all spatial derivatives of the probability distribution $P(\mathbf{x})$. Furthermore, $\mathbf{f(x)}$ cannot have singularities and has to be finite. These are reasonable conditions for biological systems as they must maintain homeostasis to survive, which ensures that the system does not enter an aberrant configuration. Altogether, the above assumptions imply that  probability distribution does not change in time:
$\frac{\partial P(\mathbf{x},t)}{\partial t}= 0=-\nabla\cdot\mathbf{J}(\mathbf{x}) $,
where $\mathbf{J}(\mathbf{x})$ is the net probability flux at position $\mathbf{x}$ in phase space. This is the law of conservation of probability at steady state.

The trivial solution is to have no net flux anywhere in the system. In this case the system is purely stochastic and deterministic dynamics play no role. Such systems are well modelled by statistical models which account only for the probabilities of different parameter configurations\cite{Tkacik2013,schneidman2006weak}. A more interesting solution exists when the deterministic dynamics form cyclic fluxes:
$\mathbf{J}(\mathbf{x}) = \nabla \times \mathbf{A}$,
where $\mathbf{A}$ is an arbitrary vector field. Cylic fluxes allow non-zero flux, and thus deterministic dynamics, by always forming closed orbits. These closed loops ensure that any probability density leaving a neighborhood of space due to the flux will be replaced by an equal amount of probability density entering the neighborhood -- and thus ensuring that the probability density is conserved. 

Cyclic fluxes help to satisfy the condition that coherent trajectories must flow along the trajectory by creating loops in the dynamics. If all points along the cyclic flux stabilize noise, then a coherent loop will be formed. Loops do not need to be independent. When two loops pass near each other in phase space it is possible for noise, or an external stimuli, to drive the system to switch loops. In this way it is even possible for one loop to diverge into two as long as the flux in and out are equal. Note that this would not be possible in a system without noise. Coherent trajectories can also decay to fixed points. This tends to occur when an external stimulus forces the system out of equilibrium and it decays back to a fixed point via a coherent trajectory. Note that when modeling the combined system this decay can still be considered a loop in the combined system. Finally, coherent trajectories can also destabilize. Destabilization can be useful when the system needs to exhibit stochastic properties. The amount of variability can be modulated by surrounding the unstable region with coherent trajectories that will “catch” the destabilized system.


Coherent trajectories can form complex systems of interconnected loops with varying degrees of stability to make use of the inherent noise in the system while still allowing for the deterministic storage of information. We suspect that such structures are required in biological systems that must reconcile their stochastic nature with the need for information storage, timing and consistency in response. By making use of the assumptions laid out above we developed a novel method for extracting these coherent trajectories which allows the simultaneous construction of a quantitative model of the system and a qualitative description of the dynamics.

\clearpage

\begin{figure}
	\centering
	\includegraphics[width=\textwidth]{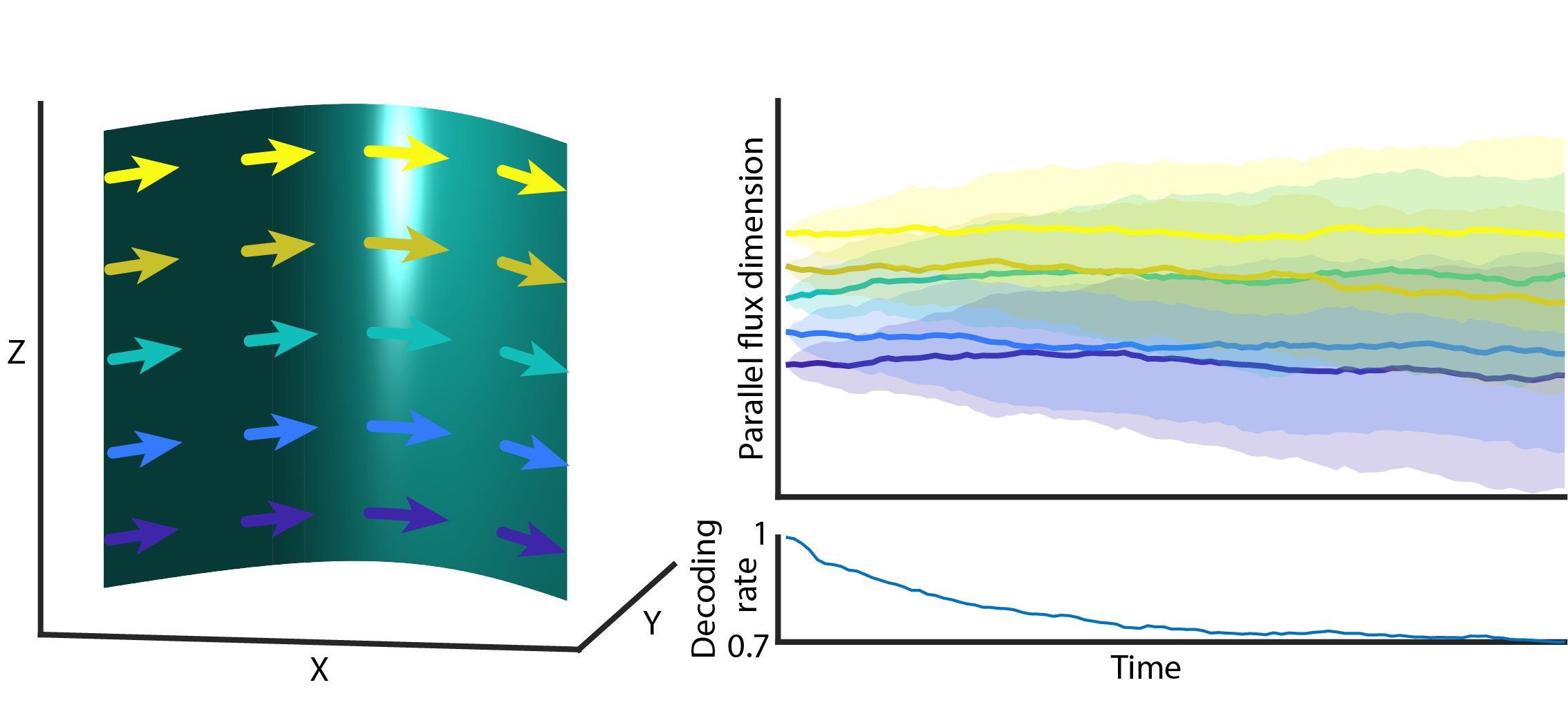}
	\phantomcaption
\end{figure}
\begin{figure}
	\ContinuedFloat
	\caption{{\textbf{Effect of Noise on 2 Dimensional Manifold.}
			\textbf{Left)} A 2D manifold such that along the Z dimension the flux is parallel. Colored arrows show the flux at different parts of the manifold. Note that this manifold can be of arbitrary non-linear shape as long as there is at least one dimension of parallel flux. \textbf{Right)} Simulations of trajectories projected onto the dimension of parallel flux. Parallel flux implies that the energy landscape along the dimension is flat. Thick lines show time-averaged values of the 50 simulations starting at each of the 5 start positions. Shaded regions are 1 standard deviation. As time passes the trajectories merge and the ability to decode the original state point decreases (Bottom left). The loss in decoding is due to the noise. If the manifold had distinct energy wells for each of the five trajectories the information would be much more stable -- but then the 2D manifold would be well approximated by five distinct 1D manifolds. Decoding is considered successful if the original start positions can be inferred from the distribution of points. Thus, for each dark blue point we count the number of non-dark blue points that are greater than the current point as correctly decoded. The same process is repeated for each start position (decoding either greater than or less than the current point). Decoding rate is the average number of correct decodings across all traces for each time point. Note that chance decoding rate is 0.5. However, the monkey has a task accuracy of 95\%. It takes only a few time steps for the decoding rate to fall below this threshold. }
		\label{Fig:noise}}
\end{figure}

\clearpage

\begin{figure}
	\centering
	\includegraphics[width=\textwidth]{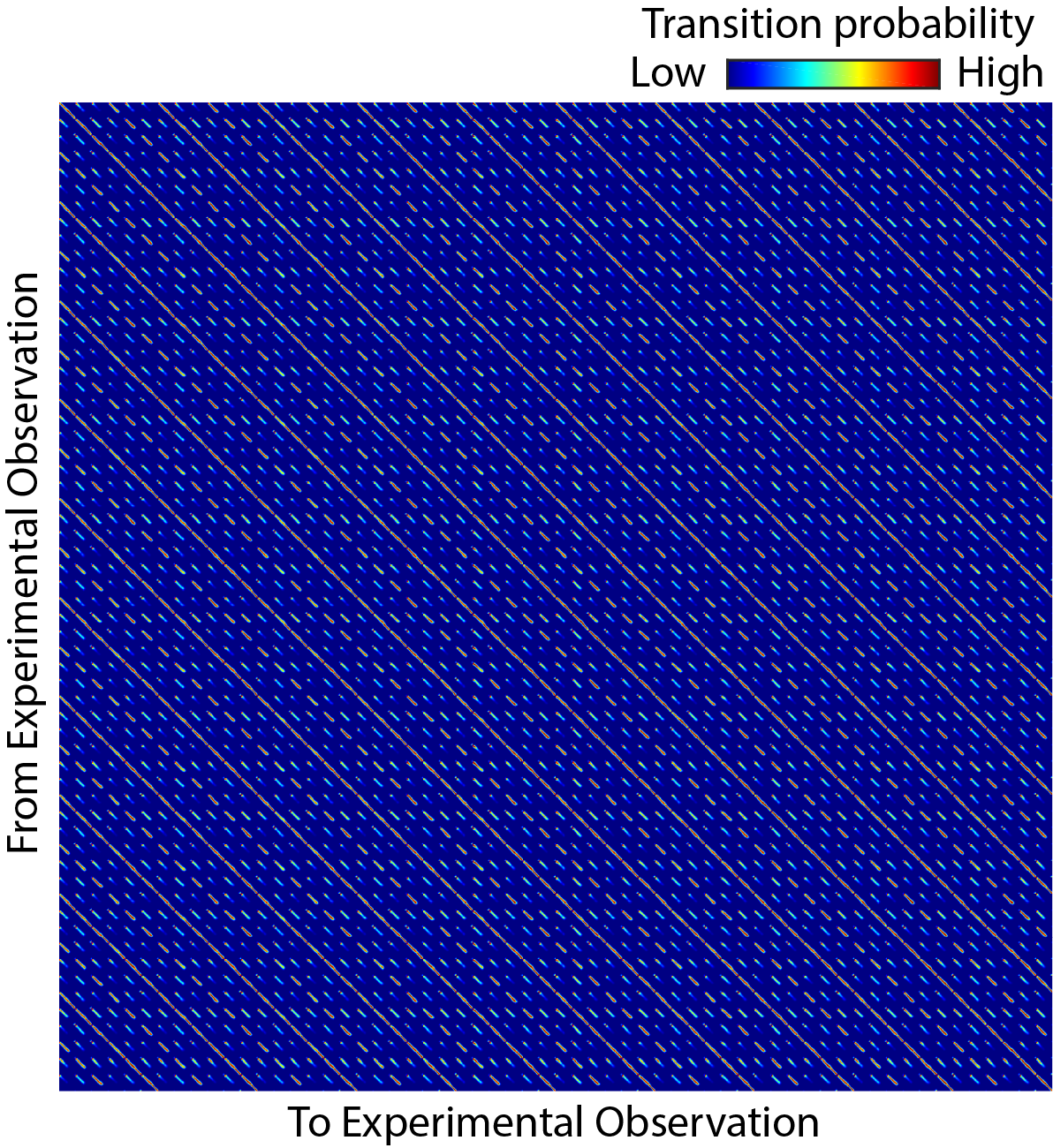}
	\phantomcaption
\end{figure}
\begin{figure}
	\ContinuedFloat
	\caption{{\textbf{Diffusion Map of the Primate Data.}
			Diffusion map constructed from the primate prefrontal cortex data in Figure~\ref{Fig:3}. Each row is the probability that a given experimentally observed point will transition to each other experimentally observed point. Note that there are many parallel lines. These represent the similarities in states across trials. There are also dashed parallel lines which correspond to the similar across conditions diverging after some time.}
		\label{Fig:diffusionMap}}
\end{figure}

\clearpage

\begin{figure}
	\centering
	\includegraphics[width=\textwidth]{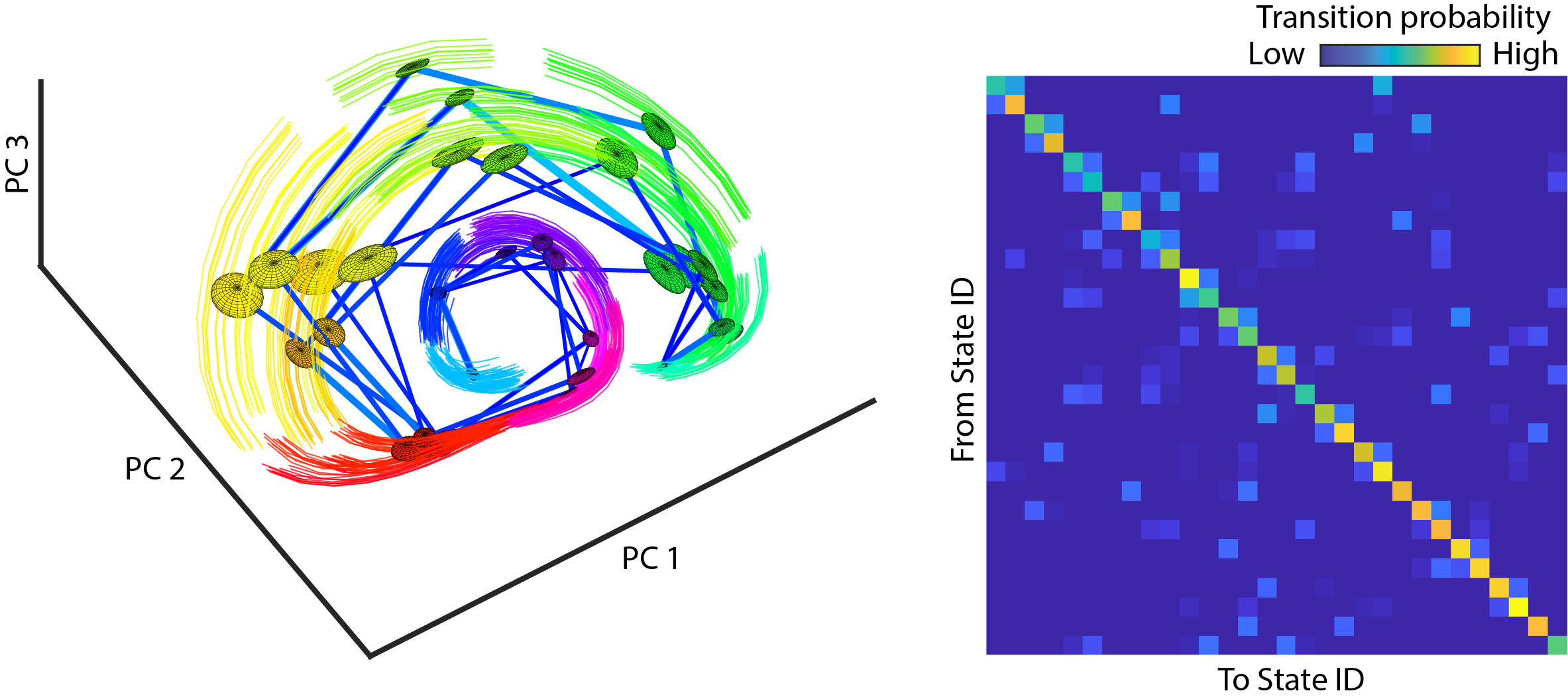}
	\phantomcaption
\end{figure}
\begin{figure}
	\ContinuedFloat
	\caption{{\textbf{Reduced Diffusion Map of the Primate Data.}
			\textbf{Left)} Mean and standard deviation of point clusters are shown as balls. Blue lines denote transitions between clusters with more than 10\% probability (Thicker and lighter colors denotes higher transition probability). Clusters are colored by mean time since trial onset. The points corresponding to each cluster are plotted as trajectories with the same color as the ball. \textbf{Right)} Transition probability matrix of the clustered matrix. Each row is the probability that a point in the given cluster will transition to a point in each other cluster. Note that all parallel lines seen in Figure~\ref{Fig:diffusionMap} have been merged together to form trajectory bundles.}
		\label{Fig:clusteredMap}}
\end{figure}

\clearpage

\begin{figure}
	\centering
	\includegraphics[width=\textwidth]{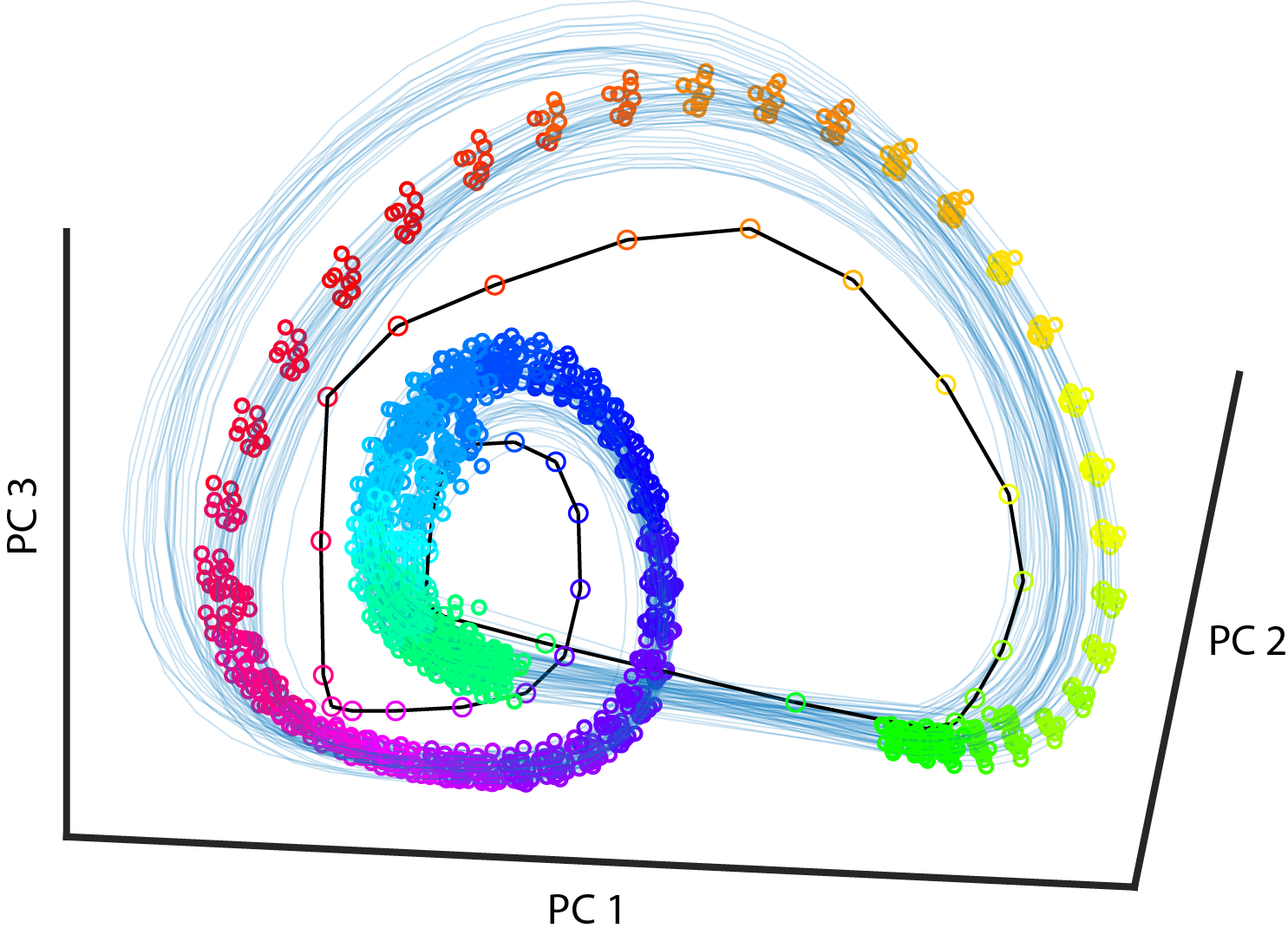}
	\phantomcaption
\end{figure}
\begin{figure}
	\ContinuedFloat
	\caption{{\textbf{Loop Clustering of Primate Data.}
			The black line denotes the first approximate loop. Each circle on this line denotes a bin, $\bar{\mathbf{x}}_{\theta(i)}$, colored by its associated phase. Note that this Gaussian smoothed loop consistently underestimates the curvature of the data. To form the final loop, points are assigned a phase value based on proximity to the approximate bins (colored points). Only points belonging to this loop are plotted. }
		\label{Fig:loopClusters}}
\end{figure}

\clearpage

\begin{figure}
	\centering
	\includegraphics[width=\textwidth]{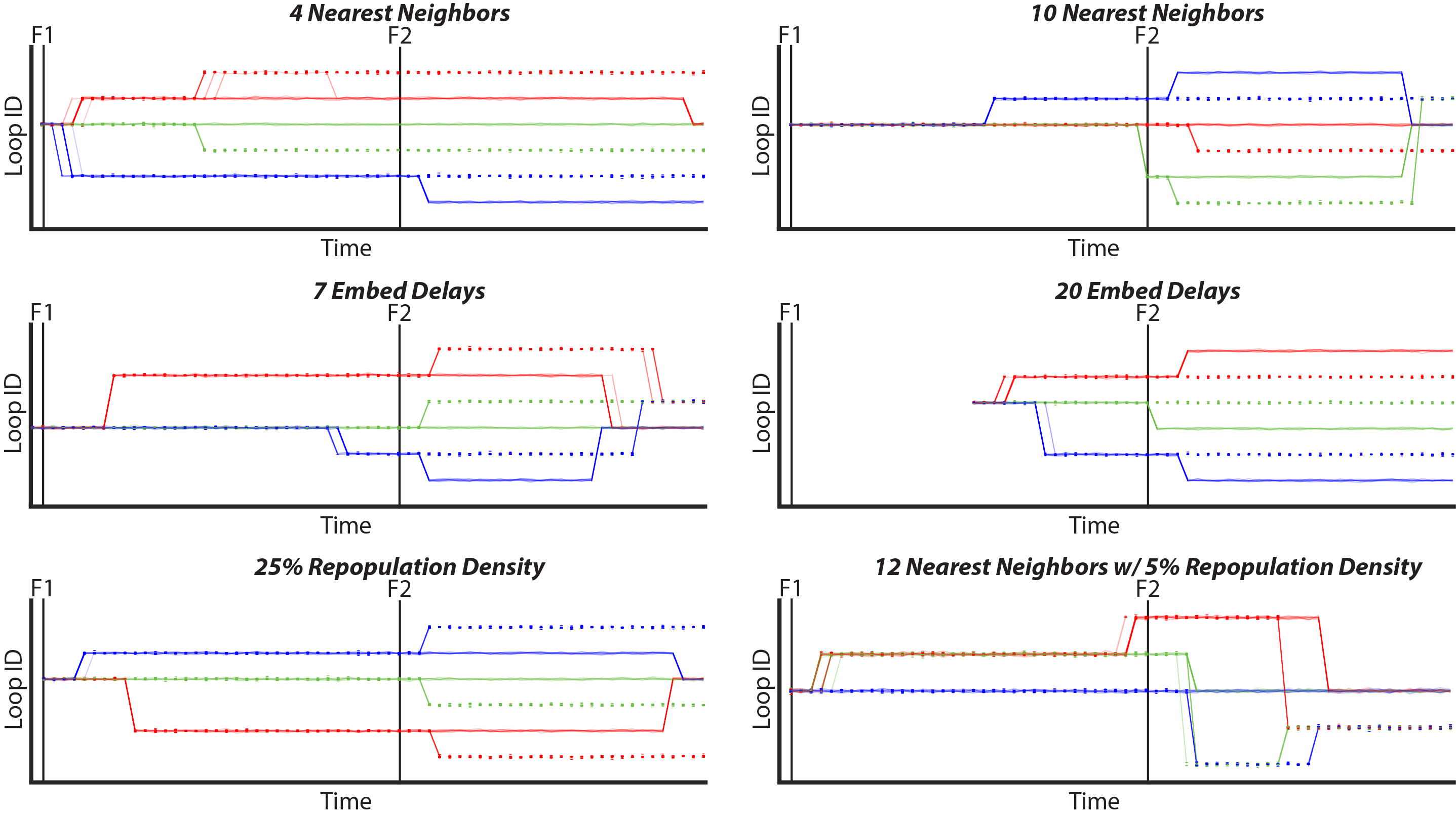}
	\phantomcaption
\end{figure}
\begin{figure}
	\ContinuedFloat
	\caption{{\textbf{Range of LOOPER parameters.}
			The LOOPER model resulting from rerunning the method on the same data from Figure~\ref{Fig:3} with small adjustments to the parameters. \textbf{Top row)} Nearest neighbors must be set in a range that is roughly the number of times each trial repeats (10 in this case). If this number is too small the trajectories will not cluster well \textbf{(left)}, if too large then all trajectories will merge together. \textbf{Middle row)} The number of delay embeds must be large enough to uncover phase space. If too few then trajectories are stickier than expected \textbf{(left)}. Note that with even 20 delays the topology looks very similar to the main text. The only issue that occurs when delay embedding is that the amount of data you have access too decreases. \textbf{Bottom row)} The method is generally robust to the repopulation density. Reducing this number helps to improve the temporal precision of the switches, at the cost of reducing loop clustering. \textbf{Right)} Adjusting the nearest neighbor count (size of local neighborhood) and the repopulation density (size of global neighborhood) can lead to some fine tuning of the diffusion map. Here the overpopulation at the local level is counteracted by the low repopulation density to produce a topology that has very different timings, but the correct overall structure. Note that the splitting after F2 is categorized by button press in this case, not distinct trajectories for each condition. }
		\label{Fig:parameterBreakdown}}
\end{figure}

\clearpage

\begin{figure}
	\centering
	\includegraphics[width=\textwidth]{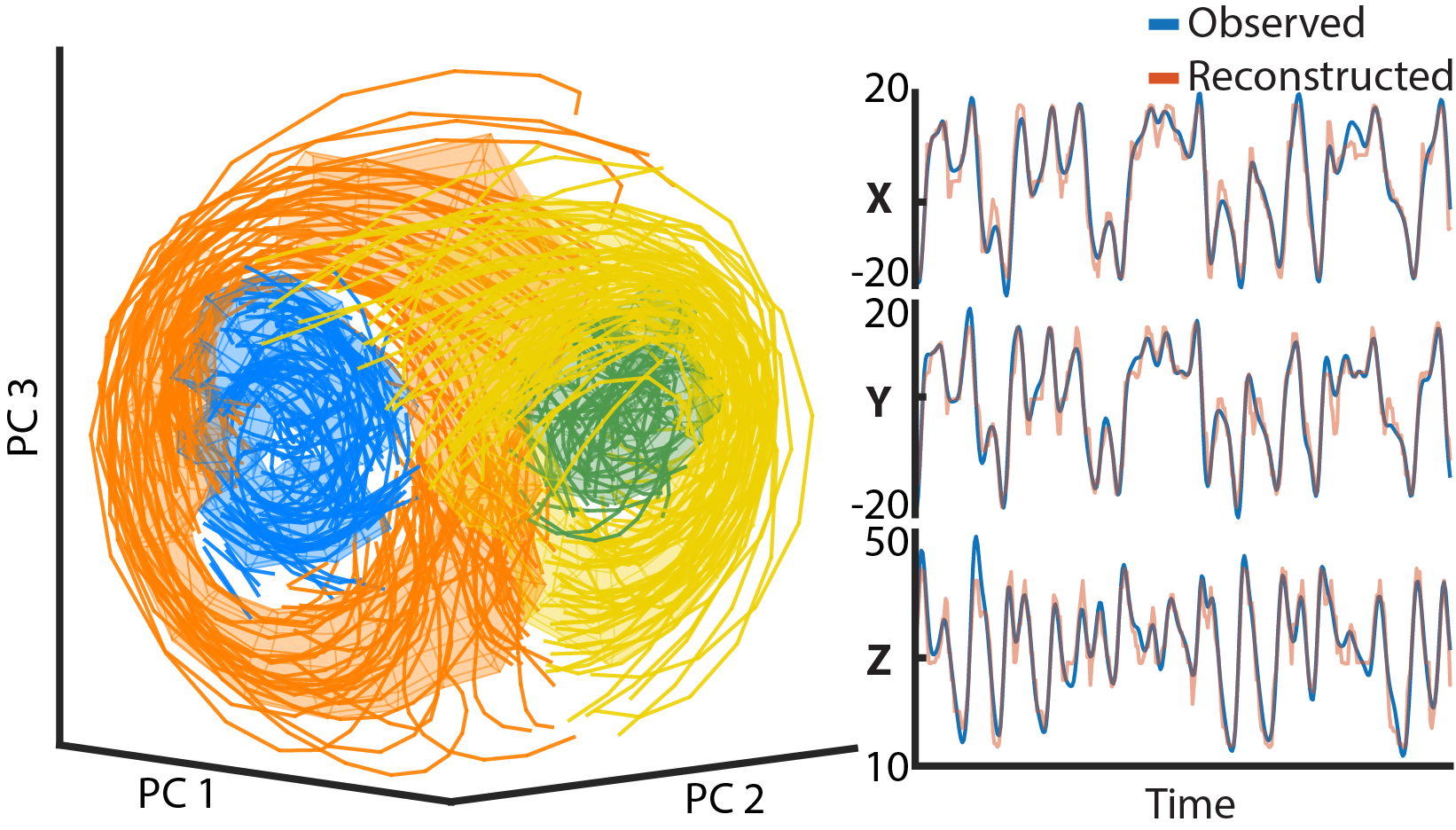}
	\phantomcaption
\end{figure}
\begin{figure}
	\ContinuedFloat
	\caption{{\textbf{LOOPER Manifold of Lorenz system.}
			\textbf{Left)} Simulation of Lorenz system with Gaussian noise (sigma=1) added at each time-step. Colored by extracted LOOPER manifold. \textbf{Right)} LOOPER reconstruction of the Lorenz system. Mean across-variable correlation is 0.96.}
		\label{Fig:Lorenz}}
\end{figure}

\clearpage

\begin{figure}
	\centering
	\includegraphics[width=\textwidth]{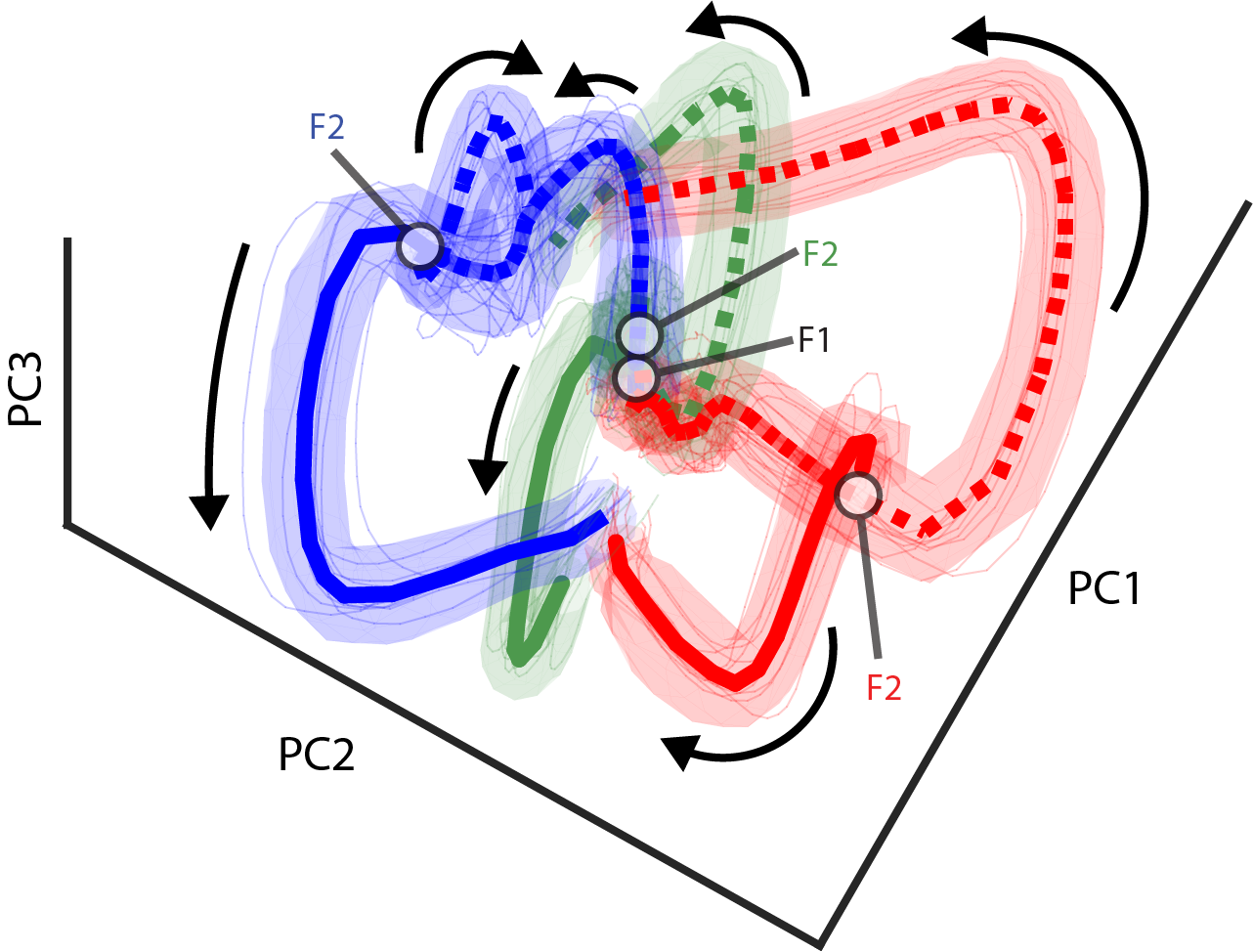}
	\phantomcaption
\end{figure}
\begin{figure}
	\ContinuedFloat
	\caption{{\textbf{Geometry of LOOPER Manifold.}
			Manifold of dynamics extracted from spiking neurons in primate prefrontal cortex as in Figure~\ref{Fig:4}D. The manifold was projected onto its demixed principal components\cite{Kobak2016}. The first two F1 stimulus dPCs and the first two F1 vs F2 comparison dPCs are used. To further increase the visibility of the distinct loop IDs, the data is delay embedded 22 times with a delay of 1 time step. Note that the only difference in this manifold from the one in the main text is the projection of the data. Timing of F1 and F2 stimuli are marked by white dots. Black arrows show direction of phase velocity around the manifold.}
		\label{Fig:geometry}}
\end{figure}

\clearpage

\begin{figure}
	\centering
	\includegraphics[width=\textwidth]{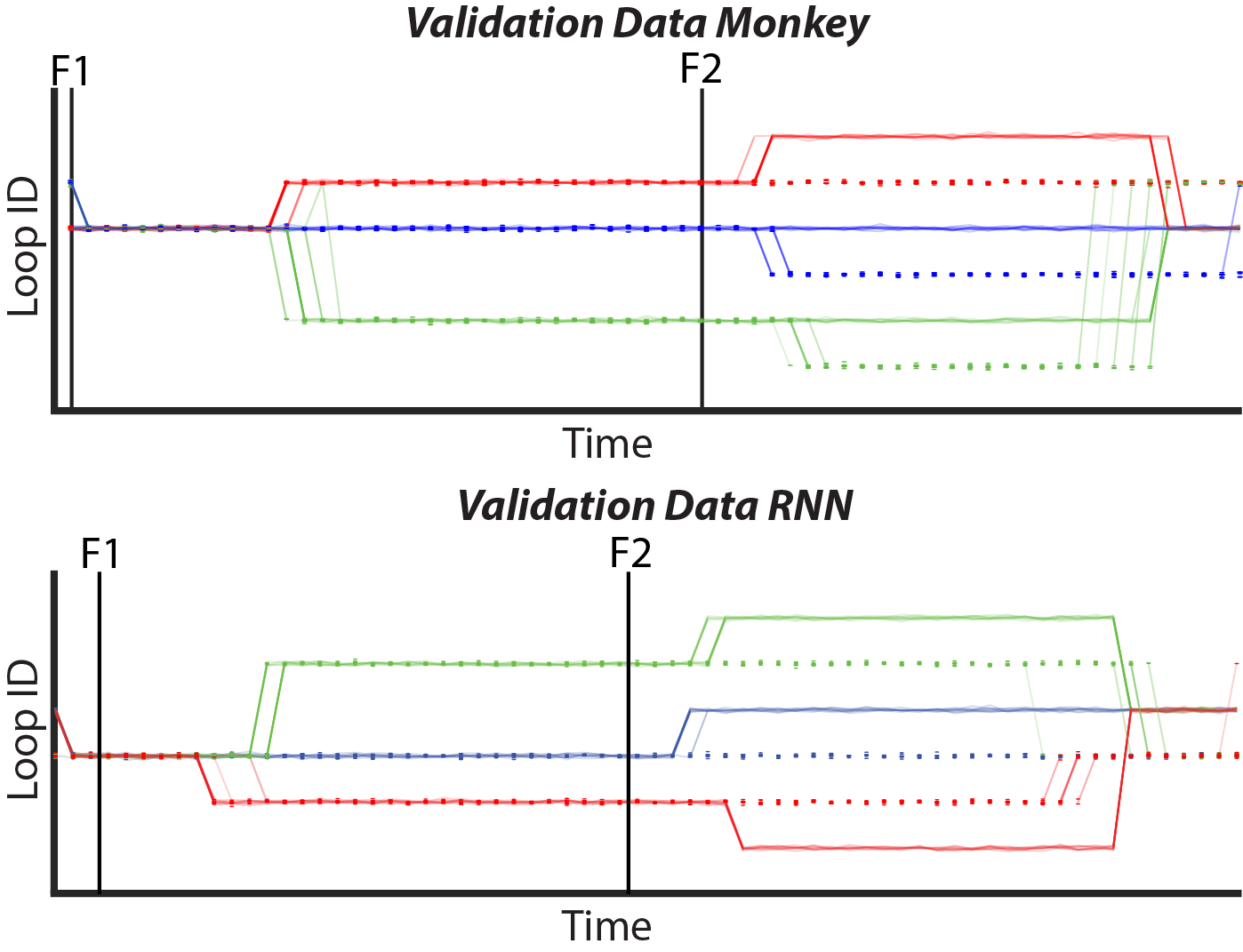}
	\phantomcaption
\end{figure}
\begin{figure}
	\ContinuedFloat
	\caption{{\textbf{Validation of LOOPER Manifolds.}
			Validation of LOOPER manifold formed by projecting the left out data from the monkey and a unique set of data from the RNN onto the original manifold as described in the Methods section. Primate validation accuracies and p-values are listed in the main text. RNN accuracies are: classifying F1 $\rightarrow$ accuracy 94\%, p-value $<$ 0.0001, remembering F1 $\rightarrow$ validation accuracy 86\%, p-value $<$ 0.0001, comparing F2 to the remembered F1 $\rightarrow$ validation accuracy 98\%, p-value $<$ 0.0001, and over all accuracy 95\%, p-value $<$ 0.0001.}
		\label{Fig:validation}}
\end{figure}

\clearpage

\begin{figure}
	\centering
	\includegraphics[width=\textwidth]{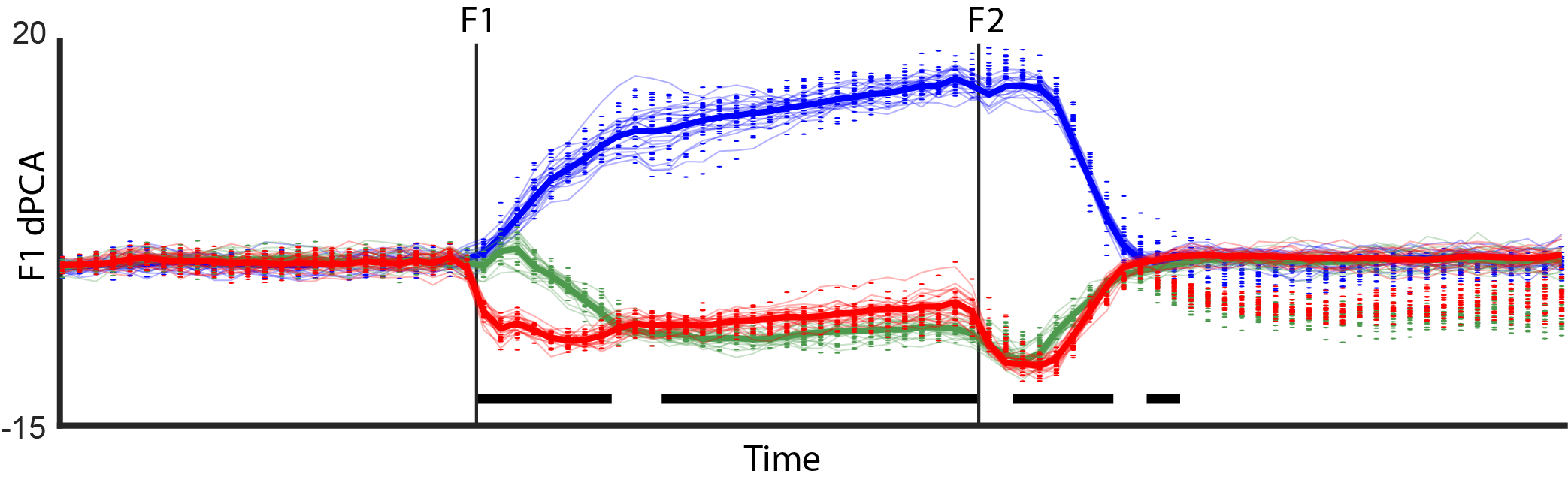}
	\phantomcaption
\end{figure}
\begin{figure}
	\ContinuedFloat
	\caption{{\textbf{Supervised learning techniques extract erroneous scaffold.}
			Activity of the poorly-conditioned RNN (Figure \ref{Fig:4}C) projected onto the dPCA which maximizes differences in the F1 condition. At each time point, the Kruskal-Wallis one-way Anova test is used to calculate a probability that the data from the F1 = 20 (green points) and F1 = 40 (red points) are drawn from the same distribution. The thick black line at the bottom shows all time points in which this test is statistically significant (p $<$ 0.01). Note that the two F1 conditions are statistically different at the onset of F2. Empirical testing shows that the response of the system at this point is not dependent on the value of F1.}
		\label{Fig:dpcaTopology}}
\end{figure}

\clearpage

\bibliography{MainTextBib}

\begin{thebibliography}{10}
\expandafter\ifx\csname url\endcsname\relax
  \def\url#1{\texttt{#1}}\fi
\expandafter\ifx\csname urlprefix\endcsname\relax\def\urlprefix{URL }\fi
\providecommand{\bibinfo}[2]{#2}
\providecommand{\eprint}[2][]{\url{#2}}

\bibitem{Afshar2011}
\bibinfo{author}{Afshar, A.} \emph{et~al.}
\newblock \bibinfo{title}{{Single-trial neural correlates of arm movement
  preparation}}.
\newblock \emph{\bibinfo{journal}{Neuron}} \textbf{\bibinfo{volume}{71}},
  \bibinfo{pages}{555--564} (\bibinfo{year}{2011}).

\bibitem{Carnevale2015}
\bibinfo{author}{Carnevale, F.}, \bibinfo{author}{DeLafuente, V.},
  \bibinfo{author}{Romo, R.}, \bibinfo{author}{Barak, O.} \&
  \bibinfo{author}{Parga, N.}
\newblock \bibinfo{title}{{Dynamic Control of Response Criterion in Premotor
  Cortex during Perceptual Detection under Temporal Uncertainty}}.
\newblock \emph{\bibinfo{journal}{Neuron}} \textbf{\bibinfo{volume}{86}},
  \bibinfo{pages}{1067--1077} (\bibinfo{year}{2015}).

\bibitem{Laurent2001}
\bibinfo{author}{Laurent, G.} \emph{et~al.}
\newblock \bibinfo{title}{{Odor Encoding as an Active, Dynamical Process:
  Experiments, Computation, and Theory}}.
\newblock \emph{\bibinfo{journal}{Annual Review of Neuroscience}}
  \textbf{\bibinfo{volume}{24}}, \bibinfo{pages}{263--297}
  (\bibinfo{year}{2001}).

\bibitem{churchland2012neural}
\bibinfo{author}{Churchland, M.~M.} \emph{et~al.}
\newblock \bibinfo{title}{{Neural population dynamics during reaching}}.
\newblock \emph{\bibinfo{journal}{Nature}} \textbf{\bibinfo{volume}{487}},
  \bibinfo{pages}{51--56} (\bibinfo{year}{2012}).
\newblock \eprint{NIHMS150003}.

\bibitem{Harvey2012}
\bibinfo{author}{Harvey, C.~D.}, \bibinfo{author}{Coen, P.} \&
  \bibinfo{author}{Tank, D.~W.}
\newblock \bibinfo{title}{{Choice-specific sequences in parietal cortex during
  a virtual-navigation decision task}}.
\newblock \emph{\bibinfo{journal}{Nature}} \textbf{\bibinfo{volume}{484}},
  \bibinfo{pages}{62--68} (\bibinfo{year}{2012}).

\bibitem{Kaufman2014}
\bibinfo{author}{Kaufman, M.~T.}, \bibinfo{author}{Churchland, M.~M.},
  \bibinfo{author}{Ryu, S.~I.} \& \bibinfo{author}{Shenoy, K.~V.}
\newblock \bibinfo{title}{{Cortical activity in the null space: Permitting
  preparation without movement}}.
\newblock \emph{\bibinfo{journal}{Nature Neuroscience}}
  \textbf{\bibinfo{volume}{17}}, \bibinfo{pages}{440--448}
  (\bibinfo{year}{2014}).

\bibitem{Kobak2016}
\bibinfo{author}{Kobak, D.} \emph{et~al.}
\newblock \bibinfo{title}{{Demixed principal component analysis of neural
  population data}}.
\newblock \emph{\bibinfo{journal}{eLife}} \textbf{\bibinfo{volume}{5}},
  \bibinfo{pages}{e10989} (\bibinfo{year}{2016}).

\bibitem{Mante2013}
\bibinfo{author}{Mante, V.}, \bibinfo{author}{Sussillo, D.},
  \bibinfo{author}{Shenoy, K.~V.} \& \bibinfo{author}{Newsome, W.~T.}
\newblock \bibinfo{title}{{Context-dependent computation by recurrent dynamics
  in prefrontal cortex}}.
\newblock \emph{\bibinfo{journal}{Nature}} \textbf{\bibinfo{volume}{503}},
  \bibinfo{pages}{78--84} (\bibinfo{year}{2013}).
\newblock \eprint{15334406}.

\bibitem{Pandarinath2015}
\bibinfo{author}{Pandarinath, C.} \emph{et~al.}
\newblock \bibinfo{title}{{Neural population dynamics in human motor cortex
  during movements in people with ALS}}.
\newblock \emph{\bibinfo{journal}{eLife}} \textbf{\bibinfo{volume}{4}},
  \bibinfo{pages}{e07436} (\bibinfo{year}{2015}).

\bibitem{Sadtler2014}
\bibinfo{author}{Sadtler, P.~T.} \emph{et~al.}
\newblock \bibinfo{title}{{Neural constraints on learning}}.
\newblock \emph{\bibinfo{journal}{Nature}} \textbf{\bibinfo{volume}{512}},
  \bibinfo{pages}{423--426} (\bibinfo{year}{2014}).

\bibitem{Gao2017}
\bibinfo{author}{Gao, P.} \emph{et~al.}
\newblock \bibinfo{title}{{A theory of multineuronal dimensionality, dynamics
  and measurement}}.
\newblock \emph{\bibinfo{journal}{bioRxiv}} \bibinfo{pages}{214262}
  (\bibinfo{year}{2017}).
\newblock \eprint{bioRxiv 214262}.

\bibitem{Pandarinath2018}
\bibinfo{author}{Pandarinath, C.} \emph{et~al.}
\newblock \bibinfo{title}{{Inferring single-trial neural population dynamics
  using sequential auto-encoders}}.
\newblock \emph{\bibinfo{journal}{Nature Methods}}
  \textbf{\bibinfo{volume}{15}}, \bibinfo{pages}{805--815}
  (\bibinfo{year}{2018}).

\bibitem{Kato2015}
\bibinfo{author}{Kato, S.} \emph{et~al.}
\newblock \bibinfo{title}{{Global Brain Dynamics Embed the Motor Command
  Sequence of Caenorhabditis elegans}}.
\newblock \emph{\bibinfo{journal}{Cell}} \textbf{\bibinfo{volume}{163}},
  \bibinfo{pages}{656--669} (\bibinfo{year}{2015}).

\bibitem{Williams2020}
\bibinfo{author}{Williams, A.~H.} \emph{et~al.}
\newblock \bibinfo{title}{{Discovering Precise Temporal Patterns in Large-Scale
  Neural Recordings through Robust and Interpretable Time Warping}}.
\newblock \emph{\bibinfo{journal}{Neuron}} \textbf{\bibinfo{volume}{105}},
  \bibinfo{pages}{246--259.e8} (\bibinfo{year}{2020}).

\bibitem{Gao2015}
\bibinfo{author}{Gao, P.} \& \bibinfo{author}{Ganguli, S.}
\newblock \bibinfo{title}{{On simplicity and complexity in the brave new world
  of large-scale neuroscience}} (\bibinfo{year}{2015}).
\newblock \eprint{1503.08779}.

\bibitem{Simonyan2015}
\bibinfo{author}{Simonyan, K.} \& \bibinfo{author}{Zisserman, A.}
\newblock \bibinfo{title}{{Very deep convolutional networks for large-scale
  image recognition}}.
\newblock In \emph{\bibinfo{booktitle}{3rd International Conference on Learning
  Representations, ICLR 2015 - Conference Track Proceedings}}
  (\bibinfo{publisher}{International Conference on Learning Representations,
  ICLR}, \bibinfo{year}{2015}).
\newblock \eprint{1409.1556}.

\bibitem{Vinyals2019}
\bibinfo{author}{Vinyals, O.} \emph{et~al.}
\newblock \bibinfo{title}{{Grandmaster level in StarCraft II using multi-agent
  reinforcement learning}}.
\newblock \emph{\bibinfo{journal}{Nature}} \textbf{\bibinfo{volume}{575}},
  \bibinfo{pages}{350--354} (\bibinfo{year}{2019}).

\bibitem{Senior2020}
\bibinfo{author}{Senior, A.~W.} \emph{et~al.}
\newblock \bibinfo{title}{{Improved protein structure prediction using
  potentials from deep learning}}.
\newblock \emph{\bibinfo{journal}{Nature}} \textbf{\bibinfo{volume}{577}},
  \bibinfo{pages}{706--710} (\bibinfo{year}{2020}).

\bibitem{Barak2017}
\bibinfo{author}{Barak, O.}
\newblock \bibinfo{title}{{Recurrent neural networks as versatile tools of
  neuroscience research}} (\bibinfo{year}{2017}).

\bibitem{Cueva2018}
\bibinfo{author}{Cueva, C.~J.} \& \bibinfo{author}{Wei, X.-X.~X.}
\newblock \bibinfo{title}{{Emergence of grid-like representations by training
  recurrent neural networks to perform spatial localization}}.
\newblock In \emph{\bibinfo{booktitle}{6th International Conference on Learning
  Representations, ICLR 2018 - Conference Track Proceedings}}
  (\bibinfo{publisher}{International Conference on Learning Representations,
  ICLR}, \bibinfo{year}{2018}).
\newblock \eprint{1803.07770}.

\bibitem{Mark}
\bibinfo{author}{Mark, S.}, \bibinfo{author}{Moran, R.}, \bibinfo{author}{Parr,
  T.}, \bibinfo{author}{Kennerley, S.} \& \bibinfo{author}{Behrens, T.}
\newblock \bibinfo{title}{{Transferring structural knowledge across cognitive
  maps in humans and models}}.
\newblock \emph{\bibinfo{journal}{bioRxiv}} \bibinfo{pages}{860478}
  (\bibinfo{year}{2019}).

\bibitem{Yamins2016}
\bibinfo{author}{Yamins, D. L.~K.} \& \bibinfo{author}{DiCarlo, J.~J.}
\newblock \bibinfo{title}{{Using goal-driven deep learning models to understand
  sensory cortex}}.
\newblock \emph{\bibinfo{journal}{Nature Neuroscience}}
  \textbf{\bibinfo{volume}{19}}, \bibinfo{pages}{356--365}
  (\bibinfo{year}{2016}).
\newblock \eprint{science.aab3050}.

\bibitem{Banino2018}
\bibinfo{author}{Banino, A.} \emph{et~al.}
\newblock \bibinfo{title}{{Vector-based navigation using grid-like
  representations in artificial agents}}.
\newblock \emph{\bibinfo{journal}{Nature}} \textbf{\bibinfo{volume}{557}},
  \bibinfo{pages}{429--433} (\bibinfo{year}{2018}).

\bibitem{Maheswaranathan2019}
\bibinfo{author}{Maheswaranathan, N.}, \bibinfo{author}{Williams, A.~H.},
  \bibinfo{author}{Golub, M.~D.}, \bibinfo{author}{Ganguli, S.} \&
  \bibinfo{author}{Sussillo, D.}
\newblock \bibinfo{title}{{Universality and individuality in neural dynamics
  across large populations of recurrent networks}}.
\newblock \bibinfo{type}{Tech. Rep.} (\bibinfo{year}{2019}).
\newblock \eprint{1907.08549}.

\bibitem{Brennan2017}
\bibinfo{author}{Brennan, C.} \& \bibinfo{author}{Proekt, A.}
\newblock \bibinfo{title}{{Universality of macroscopic neuronal dynamics in
  Caenorhabditis elegans}}  (\bibinfo{year}{2017}).
\newblock \eprint{1711.08533}.

\bibitem{Brennan2019a}
\bibinfo{author}{Brennan, C.} \& \bibinfo{author}{Proekt, A.}
\newblock \bibinfo{title}{{A quantitative model of conserved macroscopic
  dynamics predicts future motor commands}}.
\newblock \emph{\bibinfo{journal}{eLife}} \textbf{\bibinfo{volume}{8}}
  (\bibinfo{year}{2019}).

\bibitem{Raghu}
\bibinfo{author}{Raghu, M.}, \bibinfo{author}{Gilmer, J.},
  \bibinfo{author}{Yosinski, J.} \& \bibinfo{author}{Sohl-Dickstein, J.}
\newblock \bibinfo{title}{{SVCCA: Singular vector canonical correlation
  analysis for deep learning dynamics and interpretability}}.
\newblock \bibinfo{type}{Tech. Rep.} (\bibinfo{year}{2017}).
\newblock \eprint{1706.05806}.

\bibitem{Kornblith2019}
\bibinfo{author}{Kornblith, S.}, \bibinfo{author}{Norouzi, M.},
  \bibinfo{author}{Lee, H.} \& \bibinfo{author}{Hinton, G.}
\newblock \bibinfo{title}{{Similarity of neural network representations
  revisited}}.
\newblock \bibinfo{type}{Tech. Rep.} (\bibinfo{year}{2019}).
\newblock \eprint{1905.00414}.

\bibitem{nadler2006diffusion}
\bibinfo{author}{Nadler, B.}, \bibinfo{author}{Lafon, S.},
  \bibinfo{author}{Coifman, R.~R.} \& \bibinfo{author}{Kevrekidis, I.~G.}
\newblock \bibinfo{title}{{Diffusion maps, spectral clustering and reaction
  coordinates of dynamical systems}}.
\newblock \emph{\bibinfo{journal}{Applied and Computational Harmonic Analysis}}
  \textbf{\bibinfo{volume}{21}}, \bibinfo{pages}{113--127}
  (\bibinfo{year}{2006}).
\newblock \eprint{0503445}.

\bibitem{Yair2017}
\bibinfo{author}{Yair, O.}, \bibinfo{author}{Talmon, R.},
  \bibinfo{author}{Coifman, R.~R.} \& \bibinfo{author}{Kevrekidis, I.~G.}
\newblock \bibinfo{title}{{Reconstruction of normal forms by learning informed
  observation geometries from data}}.
\newblock \emph{\bibinfo{journal}{Proceedings of the National Academy of
  Sciences of the United States of America}} \textbf{\bibinfo{volume}{114}},
  \bibinfo{pages}{E7865--E7874} (\bibinfo{year}{2017}).

\bibitem{coifman2006diffusion}
\bibinfo{author}{Coifman, R.~R.} \& \bibinfo{author}{Lafon, S.}
\newblock \bibinfo{title}{{Diffusion maps}}.
\newblock \emph{\bibinfo{journal}{Applied and Computational Harmonic Analysis}}
  \textbf{\bibinfo{volume}{21}}, \bibinfo{pages}{5--30} (\bibinfo{year}{2006}).

\bibitem{Wang2008}
\bibinfo{author}{Wang, J.}, \bibinfo{author}{Xu, L.~L.} \&
  \bibinfo{author}{Wang, E.}
\newblock \bibinfo{title}{{Potential landscape and flux framework of
  nonequilibrium networks: Robustness, dissipation, and coherence of
  biochemical oscillations}}.
\newblock \emph{\bibinfo{journal}{Proceedings of the National Academy of
  Sciences}} \textbf{\bibinfo{volume}{105}}, \bibinfo{pages}{12271--12276}
  (\bibinfo{year}{2008}).
\newblock \eprint{arXiv:1408.1149}.

\bibitem{Romo1999}
\bibinfo{author}{Romo, R.}, \bibinfo{author}{Brody, C.~D.},
  \bibinfo{author}{Hern{\'{a}}ndez, A.} \& \bibinfo{author}{Lemus, L.}
\newblock \bibinfo{title}{{Neuronal correlates of parametric working memory in
  the prefrontal cortex}}.
\newblock \emph{\bibinfo{journal}{Nature}} \textbf{\bibinfo{volume}{399}},
  \bibinfo{pages}{470--473} (\bibinfo{year}{1999}).

\bibitem{van1996chaos}
\bibinfo{author}{{Van Vreeswijk}, C.} \& \bibinfo{author}{Sompolinsky, H.}
\newblock \bibinfo{title}{{Chaos in neuronal networks with balanced excitatory
  and inhibitory activity}}.
\newblock \emph{\bibinfo{journal}{Science}} \textbf{\bibinfo{volume}{274}},
  \bibinfo{pages}{1724--1726} (\bibinfo{year}{1996}).

\bibitem{rajan2016recurrent}
\bibinfo{author}{Rajan, K.}, \bibinfo{author}{Harvey, C.~D.} \&
  \bibinfo{author}{Tank, D.~W.}
\newblock \bibinfo{title}{{Recurrent Network Models of Sequence Generation and
  Memory}}.
\newblock \emph{\bibinfo{journal}{Neuron}} \textbf{\bibinfo{volume}{90}},
  \bibinfo{pages}{128--142} (\bibinfo{year}{2016}).

\bibitem{sussillo2015neural}
\bibinfo{author}{Sussillo, D.}, \bibinfo{author}{Churchland, M.~M.},
  \bibinfo{author}{Kaufman, M.~T.} \& \bibinfo{author}{Shenoy, K.~V.}
\newblock \bibinfo{title}{{A neural network that finds a naturalistic solution
  for the production of muscle activity}}.
\newblock \emph{\bibinfo{journal}{Nature Neuroscience}}
  \textbf{\bibinfo{volume}{18}}, \bibinfo{pages}{1025--1033}
  (\bibinfo{year}{2015}).

\bibitem{enel2016reservoir}
\bibinfo{author}{Enel, P.}, \bibinfo{author}{Procyk, E.},
  \bibinfo{author}{Quilodran, R.} \& \bibinfo{author}{Dominey, P.~F.}
\newblock \bibinfo{title}{{Reservoir Computing Properties of Neural Dynamics in
  Prefrontal Cortex}}.
\newblock \emph{\bibinfo{journal}{PLoS Computational Biology}}
  \textbf{\bibinfo{volume}{12}}, \bibinfo{pages}{e1004967}
  (\bibinfo{year}{2016}).

\bibitem{nichols2017global}
\bibinfo{author}{Nichols, A.~L.}, \bibinfo{author}{Eichler, T.},
  \bibinfo{author}{Latham, R.} \& \bibinfo{author}{Zimmer, M.}
\newblock \bibinfo{title}{{A global brain state underlies C. Elegans sleep
  behavior}}.
\newblock \emph{\bibinfo{journal}{Science}} \textbf{\bibinfo{volume}{356}},
  \bibinfo{pages}{1247--1256} (\bibinfo{year}{2017}).

\bibitem{Ahrens2013}
\bibinfo{author}{Ahrens, M.~B.}, \bibinfo{author}{Orger, M.~B.},
  \bibinfo{author}{Robson, D.~N.}, \bibinfo{author}{Li, J.~M.} \&
  \bibinfo{author}{Keller, P.~J.}
\newblock \bibinfo{title}{{Whole-brain functional imaging at cellular
  resolution using light-sheet microscopy}}.
\newblock \emph{\bibinfo{journal}{Nature Methods}}
  \textbf{\bibinfo{volume}{10}}, \bibinfo{pages}{413--420}
  (\bibinfo{year}{2013}).
\newblock \eprint{arXiv:1011.1669v3}.

\bibitem{Ahrens2012}
\bibinfo{author}{Ahrens, M.~B.} \emph{et~al.}
\newblock \bibinfo{title}{{Brain-wide neuronal dynamics during motor adaptation
  in zebrafish}}.
\newblock \emph{\bibinfo{journal}{Nature}} \textbf{\bibinfo{volume}{485}},
  \bibinfo{pages}{471--477} (\bibinfo{year}{2012}).

\bibitem{Flusberg2008}
\bibinfo{author}{Flusberg, B.~A.} \emph{et~al.}
\newblock \bibinfo{title}{{High-speed, miniaturized fluorescence microscopy in
  freely moving mice}}.
\newblock \emph{\bibinfo{journal}{Nature Methods}}
  \textbf{\bibinfo{volume}{5}}, \bibinfo{pages}{935--938}
  (\bibinfo{year}{2008}).

\bibitem{Katona2012}
\bibinfo{author}{Katona, G.} \emph{et~al.}
\newblock \bibinfo{title}{{Fast two-photon in vivo imaging with
  three-dimensional random-access scanning in large tissue volumes}}.
\newblock \emph{\bibinfo{journal}{Nature Methods}}
  \textbf{\bibinfo{volume}{9}}, \bibinfo{pages}{201--208}
  (\bibinfo{year}{2012}).

\bibitem{Holekamp2008}
\bibinfo{author}{Holekamp, T.~F.}, \bibinfo{author}{Turaga, D.} \&
  \bibinfo{author}{Holy, T.~E.}
\newblock \bibinfo{title}{{Fast Three-Dimensional Fluorescence Imaging of
  Activity in Neural Populations by Objective-Coupled Planar Illumination
  Microscopy}}.
\newblock \emph{\bibinfo{journal}{Neuron}} \textbf{\bibinfo{volume}{57}},
  \bibinfo{pages}{661--672} (\bibinfo{year}{2008}).

\bibitem{Grewe2010}
\bibinfo{author}{Grewe, B.~F.}, \bibinfo{author}{Langer, D.},
  \bibinfo{author}{Kasper, H.}, \bibinfo{author}{Kampa, B.~M.} \&
  \bibinfo{author}{Helmchen, F.}
\newblock \bibinfo{title}{{High-speed in vivo calcium imaging reveals neuronal
  network activity with near-millisecond precision}}.
\newblock \emph{\bibinfo{journal}{Nature Methods}}
  \textbf{\bibinfo{volume}{7}}, \bibinfo{pages}{399--405}
  (\bibinfo{year}{2010}).

\bibitem{Cheng2011}
\bibinfo{author}{Cheng, A.}, \bibinfo{author}{Gon{\c{c}}alves, J.~T.},
  \bibinfo{author}{Golshani, P.}, \bibinfo{author}{Arisaka, K.} \&
  \bibinfo{author}{Portera-Cailliau, C.}
\newblock \bibinfo{title}{{Simultaneous two-photon calcium imaging at different
  depths with spatiotemporal multiplexing}}.
\newblock \emph{\bibinfo{journal}{Nature Methods}}
  \textbf{\bibinfo{volume}{8}}, \bibinfo{pages}{139--142}
  (\bibinfo{year}{2011}).

\bibitem{Marre2012}
\bibinfo{author}{Marre, O.} \emph{et~al.}
\newblock \bibinfo{title}{{Mapping a complete neural population in the
  retina}}.
\newblock \emph{\bibinfo{journal}{Journal of Neuroscience}}
  \textbf{\bibinfo{volume}{32}}, \bibinfo{pages}{14859--14873}
  (\bibinfo{year}{2012}).

\bibitem{tenenbaum2000global}
\bibinfo{author}{Tenenbaum, J.~B.}, \bibinfo{author}{{De Silva}, V.} \&
  \bibinfo{author}{Langford, J.~C.}
\newblock \bibinfo{title}{{A global geometric framework for nonlinear
  dimensionality reduction}}.
\newblock \emph{\bibinfo{journal}{Science}} \textbf{\bibinfo{volume}{290}},
  \bibinfo{pages}{2319--2323} (\bibinfo{year}{2000}).

\bibitem{Roweis2000}
\bibinfo{author}{Roweis, S.~T.} \& \bibinfo{author}{Saul, L.~K.}
\newblock \bibinfo{title}{{Nonlinear dimensionality reduction by locally linear
  embedding}}.
\newblock \emph{\bibinfo{journal}{Science}} \textbf{\bibinfo{volume}{290}},
  \bibinfo{pages}{2323--2326} (\bibinfo{year}{2000}).

\bibitem{Scholkopf1998}
\bibinfo{author}{Sch{\"{o}}lkopf, B.}, \bibinfo{author}{Smola, A.} \&
  \bibinfo{author}{M{\"{u}}ller, K.~R.}
\newblock \bibinfo{title}{{Nonlinear Component Analysis as a Kernel Eigenvalue
  Problem}}.
\newblock \emph{\bibinfo{journal}{Neural Computation}}
  \textbf{\bibinfo{volume}{10}}, \bibinfo{pages}{1299--1319}
  (\bibinfo{year}{1998}).

\bibitem{Singer2009}
\bibinfo{author}{Singer, A.}, \bibinfo{author}{Erban, R.},
  \bibinfo{author}{Kevrekidis, I.~G.} \& \bibinfo{author}{Coifman, R.~R.}
\newblock \bibinfo{title}{{Detecting intrinsic slow variables in stochastic
  dynamical systems by anisotropic diffusion maps}}.
\newblock \emph{\bibinfo{journal}{Proceedings of the National Academy of
  Sciences of the United States of America}} \textbf{\bibinfo{volume}{106}},
  \bibinfo{pages}{16090--16095} (\bibinfo{year}{2009}).

\bibitem{Talmon2013}
\bibinfo{author}{Talmon, R.} \& \bibinfo{author}{Coifman, R.~R.}
\newblock \bibinfo{title}{{Empirical intrinsic geometry for nonlinear modeling
  and time series filtering}}.
\newblock \emph{\bibinfo{journal}{Proceedings of the National Academy of
  Sciences of the United States of America}} \textbf{\bibinfo{volume}{110}},
  \bibinfo{pages}{12535--12540} (\bibinfo{year}{2013}).

\bibitem{Stopfer2003}
\bibinfo{author}{Stopfer, M.}, \bibinfo{author}{Jayaraman, V.} \&
  \bibinfo{author}{Laurent, G.}
\newblock \bibinfo{title}{{Intensity versus identity coding in an olfactory
  system}}.
\newblock \emph{\bibinfo{journal}{Neuron}} \textbf{\bibinfo{volume}{39}},
  \bibinfo{pages}{991--1004} (\bibinfo{year}{2003}).

\bibitem{chaudhuri2019intrinsic}
\bibinfo{author}{Chaudhuri, R.}, \bibinfo{author}{Ger{\c{c}}ek, B.},
  \bibinfo{author}{Pandey, B.}, \bibinfo{author}{Peyrache, A.} \&
  \bibinfo{author}{Fiete, I.}
\newblock \bibinfo{title}{{The intrinsic attractor manifold and population
  dynamics of a canonical cognitive circuit across waking and sleep}}.
\newblock \emph{\bibinfo{journal}{Nature Neuroscience}}
  \textbf{\bibinfo{volume}{22}}, \bibinfo{pages}{1512--1520}
  (\bibinfo{year}{2019}).

\bibitem{poincare1905science}
\bibinfo{author}{Lodge, O.}
\newblock \emph{\bibinfo{title}{{Science and hypothesis}}}, vol.
  \bibinfo{volume}{123} (\bibinfo{publisher}{Science Press},
  \bibinfo{year}{1929}).

\bibitem{teschl2012ordinary}
\bibinfo{author}{Teschl, G.}
\newblock \emph{\bibinfo{title}{{Ordinary differential equations and Dynamical
  Systems}}}, vol. \bibinfo{volume}{140} (\bibinfo{publisher}{American
  Mathematical Soc.}, \bibinfo{year}{2004}).

\bibitem{beretta2018stochastic}
\bibinfo{author}{Beretta, A.}, \bibinfo{author}{Battistin, C.},
  \bibinfo{author}{de~Mulatier, C.}, \bibinfo{author}{Mastromatteo, I.} \&
  \bibinfo{author}{Marsili, M.}
\newblock \bibinfo{title}{{The stochastic complexity of spin models: Are
  pairwise models really simple?}}
\newblock \emph{\bibinfo{journal}{Entropy}} \textbf{\bibinfo{volume}{20}},
  \bibinfo{pages}{739} (\bibinfo{year}{2018}).
\newblock \eprint{1702.07549}.

\bibitem{dijkstra1959note}
\bibinfo{author}{Dijkstra, E.~W.}
\newblock \bibinfo{title}{{A note on two problems in connexion with graphs}}.
\newblock \emph{\bibinfo{journal}{Numerische Mathematik}}
  \textbf{\bibinfo{volume}{1}}, \bibinfo{pages}{269--271}
  (\bibinfo{year}{1959}).

\bibitem{Vilares2011}
\bibinfo{author}{Vilares, I.} \& \bibinfo{author}{Kording, K.}
\newblock \bibinfo{title}{{Bayesian models: The structure of the world,
  uncertainty, behavior, and the brain}} (\bibinfo{year}{2011}).

\bibitem{Low2018}
\bibinfo{author}{Low, R.~J.}, \bibinfo{author}{Lewallen, S.},
  \bibinfo{author}{Aronov, D.}, \bibinfo{author}{Nevers, R.} \&
  \bibinfo{author}{Tank, D.~W.}
\newblock \bibinfo{title}{{Probing variability in a cognitive map using
  manifold inference from neural dynamics}}.
\newblock \emph{\bibinfo{journal}{bioRxiv}} \bibinfo{pages}{418939}
  (\bibinfo{year}{2018}).

\bibitem{Anderson2014}
\bibinfo{author}{Anderson, A.} \emph{et~al.}
\newblock \bibinfo{title}{{Non-negative matrix factorization of multimodal MRI,
  fMRI and phenotypic data reveals differential changes in default mode
  subnetworks in ADHD}} (\bibinfo{year}{2014}).

\bibitem{Tkacik2013}
\bibinfo{author}{Tka{\v{c}}ik, G.} \emph{et~al.}
\newblock \bibinfo{title}{{The simplest maximum entropy model for collective
  behavior in a neural network}}.
\newblock \emph{\bibinfo{journal}{Journal of Statistical Mechanics: Theory and
  Experiment}} \textbf{\bibinfo{volume}{2013}}, \bibinfo{pages}{P03011}
  (\bibinfo{year}{2013}).
\newblock \eprint{1207.6319}.

\bibitem{schneidman2006weak}
\bibinfo{author}{Schneidman, E.}, \bibinfo{author}{Berry, M.~J.},
  \bibinfo{author}{Segev, R.} \& \bibinfo{author}{Bialek, W.}
\newblock \bibinfo{title}{{Weak pairwise correlations imply strongly correlated
  network states in a neural population}}.
\newblock \emph{\bibinfo{journal}{Nature}} \textbf{\bibinfo{volume}{440}},
  \bibinfo{pages}{1007--1012} (\bibinfo{year}{2006}).
\newblock \eprint{0512013}.

\end{thebibliography}

\end{document}